\documentclass[sigconf]{acmart}

\renewcommand\footnotetextcopyrightpermission[1]{} 
\renewcommand\footnotetextcopyrightpermission[1]{}

\acmConference[]{}{}

\usepackage{subcaption}
\usepackage[inline,shortlabels]{enumitem}
\usepackage{import}
\usepackage{amsmath}
\usepackage{algorithm}
\usepackage[noend]{algpseudocode}
\usepackage{graphicx}
\usepackage{enumitem}
\usepackage{multirow}
\usepackage{xspace}
\usepackage{etoolbox}
\usepackage{verbatim}
\usepackage{setspace}
\usepackage{url}
\usepackage{array}
\usepackage{mathtools}
\usepackage{colortbl}
\usepackage{booktabs}
\usepackage{enumitem}

\newcommand{\MethodName}{SPIREL\xspace}
\newcommand{\etal}{\emph{et al.} }
\makeatletter
\newcommand\newref[1]{#1\def\@currentlabel{#1}}
\newcommand\newtag[2]{(#1)\def\@currentlabel{#1}\label{#2:#1}}
\makeatother

\newcolumntype{x}[1]{>{\centering\arraybackslash\hspace{0pt}}p{#1}}
\newcolumntype{H}{>{\setbox0=\hbox\bgroup}c<{\egroup}@{}}
\begin{document}

\title[Security Policy Implementation Using DRL]{Transferable Cost-Aware Security Policy Implementation for Malware Detection Using Deep Reinforcement Learning}

\author{Yoni Birman, Shaked Hindi, Gilad Katz, Asaf Shabtai}
\affiliation{
 \institution{Department of Software and Information Systems Engineering \\ Ben-Gurion University of the Negev}
}
\email{{birman,shakedhi}@post.bgu.ac.il,  {giladkz,shabtaia}@bgu.ac.il}

\begin{abstract}
Malware detection is an ever-present challenge for all organizational gatekeepers, who must maintain high detection rates while minimizing interruptions to the organization's workflow. 
To improve detection rates, organizations often deploy an ensemble of detectors. 
While effective, this approach is computationally expensive, since every file -- even clear-cut cases -- needs to be analyzed by all detectors. 
Moreover, with an ever-increasing number of files to process, the use of ensembles may incur unacceptable processing times and costs (e.g., cloud resources).
In this study, we propose \MethodName, a reinforcement learning-based method for cost-effective malware detection.
Our method enables organizations to directly associate costs to correct/incorrect classification, computing resources and run-time, and then dynamically establishes a security policy.
This security policy is then implemented, and for each inspected file, a different set of detectors is assigned and a different detection threshold is set. 
Our evaluation on two malware domains -- Portable Executable (PE) and Android Application Package (APK) files -- shows that \MethodName is both accurate and extremely resource-efficient: the proposed method either outperforms the best performing baselines while achieving a modest improvement in efficiency, or reduces the required running time by \textasciitilde80\% while decreasing the accuracy and F1-score by only 0.5\%. 
We also show that our approach is both highly transferable across different datasets and adaptable to changes in individual detector performance.
\end{abstract}


\maketitle

\section{\label{sec:introduction}Introduction}
Organizational gatekeepers are constantly faced with two (sometimes conflicting) needs: \textit{security} and \textit{availability}. 
Maintaining security means ensuring that the organization is well protected from malicious attacks. 
The need to maintain availability means that security is achieved with minimal disruptions to the flow of information in and out of the organization. 
Since availability is a function of allocated computing power, we refer to striking the right balance as the cost/availability problem.

On the security front, one of the primary techniques used by organizations is malware detection.
The use of advanced obfuscation techniques~\cite{p29} has increasingly driven organizations to transition from signature-based malware detection~\cite{deepmalnet} to machine learning (ML)-based solutions.
ML-based approaches are considered more effective than their signature-based counterparts in detecting previously unknown malware~\cite{p20,p21,p22,p23,p25}, but they usually suffer from a relatively high false positive rate~\cite{p78}, which renders the use of a single ML-based detector insufficient on its own.

In order to improve detection rates (reduce the false positive rate) and make it harder for attackers to evade detection, it is common to deploy an ensemble of detectors. 
These ensembles often include multiple signature-based and ML-based detectors, with the output of all detectors combined using simple heuristics (e.g., voting) or more advanced ML algorithms~\cite{p95}. 
Ensembles are common in commercial products, including Microsoft Defender Advanced Threat Protection,\footnote{https://www.microsoft.com/en-us/microsoft-365/windows/microsoft-defender-atp} OPSWAT MetaDefender,\footnote{https://www.opswat.com/products/metadefender} and VirusTotal.\footnote{https://www.virustotal.com}

While ensemble approaches improve security (i.e., detection rates), they often come with a heavy price in terms of cost and availability. 
Current ensemble solutions require that each incoming file be analyzed by \textit{all} of the ensemble's detectors. 
This means that the time required to analyze each file is equal to or greater than that of the most time-consuming detector.
In addition, the large number of detectors requires expensive computing resources.
To face an ever-increasing volume of incoming files, organizations must either spend a great deal of money (on new hardware or cloud resources) or accept longer processing times and possible negative consequences to their operations. 

An ideal solution to the cost/availability problem presented above is one in which each file is analyzed by the most computationally efficient subset of detectors required to correctly classify it. 
Despite the obvious advantages of such an implementation, to the best of our knowledge, no studies have explored such an approach and no existing commercial products offer this capability. 
We argue that the reason this research direction has not been pursued is that it is \textit{incompatible with standard ML-based approaches}. 

ML algorithms require features (i.e., inputs) in order to reach a decision. 
When training a ML model to select the detectors to be applied to a given file, these features can be obtained in two ways: \textit{a)} extracting file attributes prior to running the detectors, and \textit{b)} running some of the detectors and then using their classifications as the features for selecting the next detector. 
The former approach is likely to require long running times and a significant amount of resources, since extracting even a simple set of features for every file on an enterprise scale is a far from trivial task. 
Therefore, the savings obtained by applying this approach is likely to be limited. 
The latter approach is impractical, since it would require training a \textit{separate ML model for every possible detector combination}. 
Moreover, the introduction of a new detector to the ensemble would require retraining all models.
In addition to the challenges described above, existing ML-based approaches offer no clear way of integrating the performance/cost/availability trade-offs into their target function. 
We argue that deep reinforcement learning (DRL) can address all of the abovementioned challenges.

In this study, we present \MethodName, a reinforcement learning-based framework for efficiently managing a multi-detector malware detection platform. 
For each file, our approach dynamically queries a subset of the detectors and then determines --- based on the feedback it receives --- whether enough information exists to classify the file. 
\MethodName then proceeds to either classify the file (thus ending the analysis process) or query additional detectors. 
\MethodName's decision-making process is governed by an objective function that quantifies the costs and benefits of correct and incorrect file classification, as well as the costs of the computing resources, measured in this study by the objective measure of runtime. 
This function can be easily modified to reflect different organizational priorities.

We evaluate \MethodName on large datasets from two malware domains: Portable Executable (PE) files and Android Application Package (APK) files. 
We implement and evaluate five different objective functions (i.e., organizational policies), each expressing a different preference in the security/availability trade-off.
We show that the strategies developed based on our objective functions are superior to those obtained by \textit{all possible} detector combinations \textit{regardless of the cost/availability trade-off preferred by the organization}. 
For example, \MethodName can either outperform the most-accurate baseline while reducing running time by \textasciitilde2.3\% or reduce running time by \textasciitilde80\% while reducing the accuracy and F1 scores by only 0.5\%.
Moreover, we demonstrate that our approach is both highly transferable, with models trained on one dataset easily deployed to another, and quick to adapt to changes in the performance of individual detectors due to external factors (e.g., a new form of attack that makes a detector ineffective). 
To summarize, our contributions are as follows:
\begin{itemize}
    \item We introduce \MethodName, an RL-based framework for malware detection. 
    \MethodName intelligently and dynamically assigns detector subsets to each file, offering superior detection capabilities to all evaluated baselines at a lower computational cost. 
    \MethodName is also easily configurable, enabling each organization to define its own preferences with regard to resource use and detection performance. The framework even enables organizations to define the relative cost of each error type (false positive vs. false negative). Our approach is the first to offer all of these capabilities.
    \item The framework is generic and can be applied on all types of detectors: static and dynamic, proprietary and off-the-shelf.
    It can also be applied to any file type (EXE, APK, PDF, etc.)
    \item \MethodName is highly effective in a transfer learning setting. We show that pretrained models reduce the training time for new datasets and also improve performance when the training data in the target domain is limited.
    \item We propose a method for organizations that integrates the monetary and computational costs (e.g., execution time, CPU and RAM, the cost of electricity) into their security policy. By doing so we enable organizations with different levels of security preferences and operational priorities (and budgets) to define the security policies best suited to their needs.
    \item Finally, we make our code\footnote{https://anonymous.4open.science/r/c2314b1f-185a-445d-8909-c22c1ace1633/} and datasets publicly available.
\end{itemize}

\section{RELATED WORK}

\subsection{Malware Detection in Portable Executables}

PE files can be represented in various ways, a fact that has contributed to the large number of approaches proposed for their analysis. 
The most common approach for representing a PE is by finding unique signatures~\cite{p46,park2014intelligent}. 
This method is frequently used, as matching signatures to a database is efficient and ensures zero false positives. 
Although very common, signature-based methods cannot detect unknown malware types or detecting known malware that has been even slightly changed.
Also, because signature databases are often limited, most malware remains undetected~\cite{arshad2016android}. 

More recent studies have proposed multiple enhancements to the signature-based approach. 
These enhancements include the extraction of meta-data about the analyzed files, as well as the use of ML. 
Parisotto~\etal~\cite{p34} describe the use of disassembly to generate another type of $n$-gram features called opcode $n$-grams. These features were used an input to various algorithms such as deep networks, decision trees and Naive Bayes. 
Additional studies~\cite{p22, p26, p33} propose the use of file metadata to improve detection. 
Other studies take advantage of the ordered nature of PE files. Raman~\etal~\cite{p33}, for example, used only seven features extracted from the PE headers to classify malicious files using Random Forest and Support Vector Machine algorithms for classification. 

One common approach to increase malware detection is combining multiple detectors---an ensemble. Ensembles are common in commercial products (e.g., VirusTotal and MetaDefender) and have various implementations. 
For example, the authors of~\cite{menahem2009improving} combined five individual classifiers to achieve higher detection rates with little effect on execution times. 
The use of advanced ML algorithms is also common for integrating the outputs of the various detectors~\cite{lee2019api}.

\subsection{Malware Detection in Android Packages} 

Malware detection in APKs also utilizes signature-based and ML-based methods. 
For ML-based techniques, the main difference between the two domains stems from the features that are extracted for the training of the models. 
Sato \etal~\cite{sato2013detecting}, for example, extracted textual and numeric features from the APK's manifest file to determine its malignancy. 
Another approach to analyzing the manifest file was presented by~\cite{wang2016mmda}, who also analyzed the most frequent permissions.
Dalvik bytecode frequency analysis was also studied, with Kang~\etal~\cite{kang2013android} using it to identify malware families, and Chan~\etal~\cite{chan2014static} leveraging its permissions and calls to improve malware detection.

In recent years, deep learning algorithms have proven themselves highly effective in the field of malware detection. Abderrahmane~\etal~\cite{abderrahmane2019android} used convolutional neural networks to analyze APK calls in the Linux kernel, while~\cite{vinayakumar2017deep} proposed using a long short-term memory recurrent neural network (LSTM-RNN) for a similar purpose.  Milosevic~\etal~\cite{milosevic2017machine} proposed  deep learning techniques to perform static analysis, where both permissions and source-code analysis were employed. 

As in the analysis of PEs, ensemble-based methods have proven themselves highly effective.
In \cite{yerima2015high}, the authors proposed combining the Random Forest algorithm together with logistic regression to analyze Android API calls. 
Pekta\c{s}~\etal~\cite{pektacs2017ensemble} presented a hybrid feature-based classification framework capable of analyzing both static (such as permissions and hidden payload) and dynamic (API calls, installed services and network connections) features.
\vspace*{-4pt}

\subsection{\label{subsec:deepRL}Reinforcement Learning and Security}

Reinforcement learning (RL) is an area of machine learning that addresses decision-making in complex scenarios, including those in which only partial information is available~\cite{kalweit2017uncertainty}. 
The ability of RL algorithms to explore large solution spaces and devise highly efficient strategies to address them (especially when coupled with deep learning) was shown to be highly effective in areas such as robotics and control problems~\cite{schulman2015trust}, genetic algorithms~\cite{such2017deep}, and complex games~\cite{silver2017mastering}.
RL tasks normally consist of an \textit{agent} that interacts with an \textit{environment} in a sequence of actions and rewards.
In each time step $t$, the agent selects an action $a_t$ from $A=\{a_1, a_2,...,a_k\}$ that both modifies the state of the environment and incurs a reward $r_t$, which is either positive or negative (please note that for convenience we use the term ``cost'' to describe negative rewards). 
The goal of the agent is to interact with the environment in a way that maximizes future rewards $R_t= \sum_{t}^ {T} r_t$ in time span $\{t..T\}$.

RL is used in the security domain mainly for adversarial learning and malware detection. 
In the field of malware detection, Silver~\etal~\cite{p25} presented a proof of concept for an adaptive rule-based malware detection framework that employs a learning classifier combined with a rule-based expert system.
Then, an RL algorithm was used to determine whether a PE is malicious. Another study~\cite{p80} propsed using RL to classify different malware types utilizing a set of features commonly used by anti-virus software.
A similar method was presented by Wan~\etal~\cite{p79} for optimizing malware detection accuracy on mobile devices in which the detection process was offloaded to cloud servers using an RL algorithm.

All of the studies presented above focus on improving or evading the malware detection process. 
In contrast, our proposed approach focuses on the previously unaddressed goal of increasing availability.
In other words, our aim is to select a subset of malware detectors in a way that maintains performance while reducing resource use.

\subsection{Transfer Learning}

Transfer learning (TL)~\cite{weiss2016survey} is a field in machine learning that focuses on utilizing knowledge gained in a source task/domain and using it to facilitate the learning of a target predictive function in a different target task/domain. TL can be roughly divided into two categories:
\begin{enumerate*}[(i)]
    \item \textit{homogeneous TL} refers to TL tasks where the feature and label spaces of both the source and target domains are the same size; and \item \textit{heterogeneous TL} refers to TL tasks where the feature space and/or the label space of both the source and target domains are different sizes. 
\end{enumerate*}
Homogeneous TL  is commonly referred to as a parameter-based approach with hard weight sharing~\cite{day2017survey}, and this is the approach used in our experiments (see Section \ref{subsec:transfer}).

TL has been shown to have the potential to significantly improve a reinforcement learning agent's sample efficiency~\cite{parisotto2015actor}. 
Usually, TL is performed for two reasons:
\begin{enumerate*}[(i)]
    \item \textit{there is a need for quick deployment} --- when starting with a fully-trained network, the training convergence time for the new domain is usually faster~\cite{narvekar2016source, lazaric2012transfer}; and
    \item \textit{there is a small amount of target data} --- the knowledge from the source domain can be used to effectively learn using less amount of data from the target domain~\cite{hu2016transfer, ng2015deep}.
\end{enumerate*}

In the field of security, TL has been mostly used to achieve the above mentioned goals. Chen~\etal~\cite{chen2018deep} applied TL from computer vision to static malware classification. A similar study \cite{rezende2017malicious} adapted the ResNet-50 architecture to classify malware families. Nahmias~\etal~\cite{nahmias2019trustsign} utilized the VGG-19 architecture (also trained on ImageNet) to generate malware signatures. It is important to note that our use of TL is different than previous studies in the sense that we do not aim to transfer detection capabilities, but rather the strategies that were developed to address the cost/availability problem.
\vspace*{-3pt}

\begin{figure}[bp]
  \centering
  \includegraphics[scale=0.32]{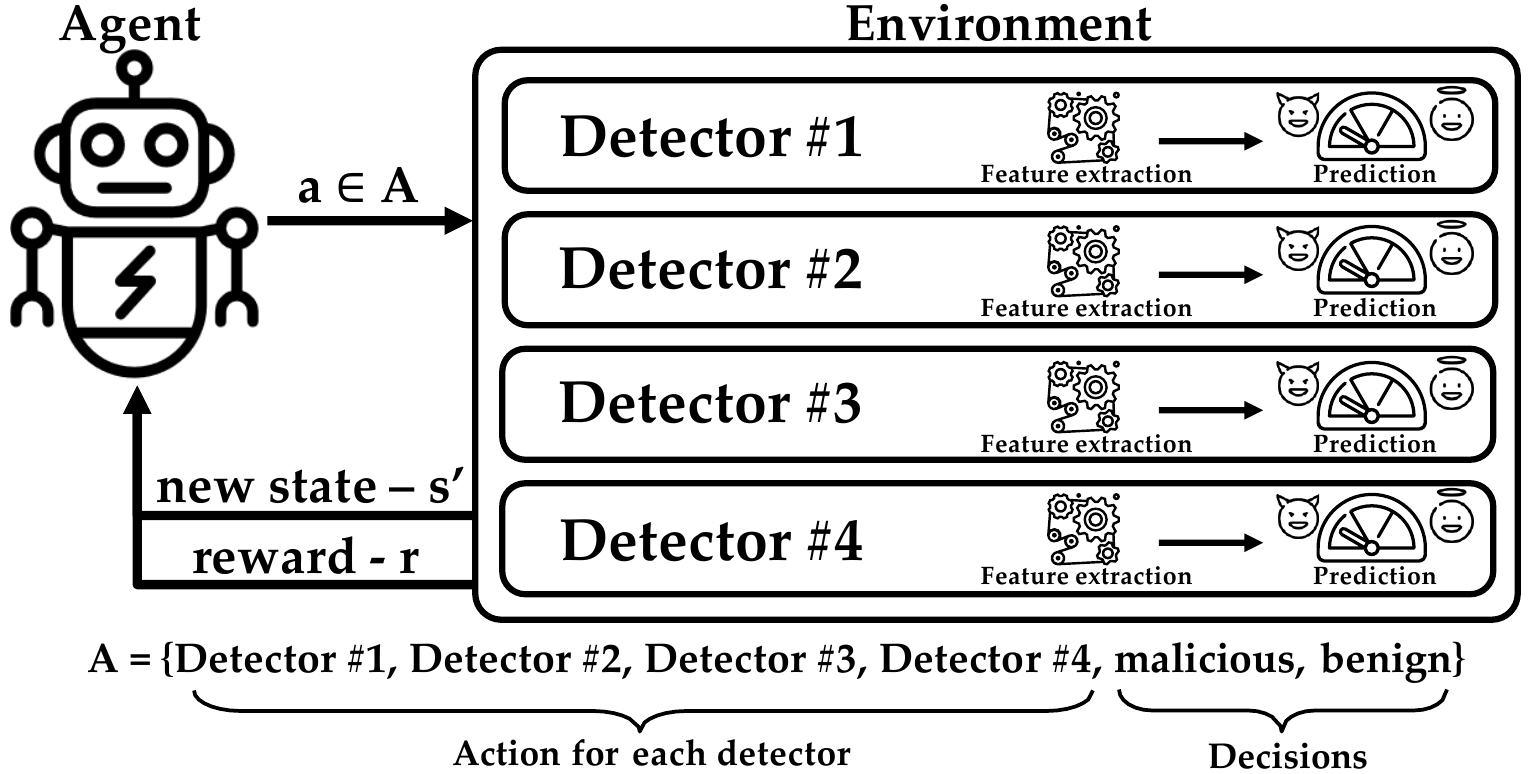}
  \caption{A high-level illustration of \MethodName's architecture. 
  The agent interacts with the environment, either choosing to submit the analyzed file for further analysis (i.e., invoking another malware detector) or terminating the process and producing a classification: malicious or benign.}
  \label{fig:rlarch}
\end{figure}

\section{Our Method}
We present \MethodName, a \textbf{s}ecurity \textbf{p}olicy \textbf{i}mplementation using deep \textbf{re}inforcement \textbf{l}earning.
The goal of our approach is to automatically ``learn'' (i.e., generate) a security policy that best fits an organization's priorities and requirements. 
More specifically, we train a deep neural network to \textit{dynamically} determine when enough information exists to classify a given file and when more analysis is needed. 
The policy generated by our approach is shaped by the rewards and penalties assigned to correct and incorrect file classification, as well as by the cost of using computing resources.
We introduce a reinforcement learning (RL) framework that explores the efficacy of various detector combinations and continuously performs cost-benefit analysis to select the optimal security policy. 

The challenge of selecting different detector combinations for each file can be modeled as an exploration/exploitation problem. 
While the cost (i.e., computing resources) of using a detector can be closely approximated in advance, its benefit --- the additional information gained by using the detector to analyze the file --- is more difficult to predict. 
RL algorithms are designed to solve exploration/exploitation problems and are therefore well suited for the task at hand. 
Moreover, DRL is known to perform well in high-uncertainty scenarios with partial information~\cite{kalweit2017uncertainty}.

\MethodName is presented in Figure~\ref{fig:rlarch}. 
In each step the agent chooses whether to further analyze the file by invoking a malware detector or to produce a classification -- either malicious or benign -- for the analyzed file. 
Once it produces a classification, the process terminates. 
As in all RL systems, our solution consists of three components: \textit{states}, \textit{actions}, and \textit{rewards}. 
Next, we describe each component in detail.

\begin{figure}[htp!]
  \centering
  \includegraphics[scale=1.4]{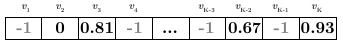}
  \vspace*{-6pt}
  \caption{An example of a state vector for $K$ malware detectors.
  The -1 values indicate that the detector was not yet activated.
  A score in the range [0,1] indicates the maliciousness score produced by an activated detector.}
  \label{fig:statevec}
  \vspace*{-6pt}
\end{figure}

\noindent \textbf{States.} In RL, the set of states (i.e., the environment) represents all possible scenarios the algorithm may encounter. 
In chess, for example, the environment is made up of all possible board positions. 
In our case, the set of states consists of all possible confidence score (i.e., certainty level) combinations of the participating detectors. 
More specifically, for a malware detection environment consisting of $K$ detectors, each possible state will be represented by a vector $V=\{v_1,v_2,\dots,v_K\}$, with the value of $v_x$ set by:

\begin{equation}
v_x =
\begin{cases*}
\,[0,1] & if detector $x$ has been applied \\
\,-1    & otherwise
\end{cases*}
\end{equation}

\noindent Therefore, the initial state for each incoming file is a vector consisting entirely of -1 values. 
As various detectors are chosen to analyze the files, entries in the vector are populated with the confidence scores they provide. 
All scores are normalized to a [0,1] range, where a confidence value of one indicates full certainty of the file being malicious, and zero indicates full certainty of its being benign. 
An example of a possible state vector is presented in Figure~\ref{fig:statevec}. 
It is also important to emphasize that \MethodName requires no additional information on the analyzed files. 
In addition to making our approach more generic and less dependent on available meta-data, \MethodName is also more computationally efficient as a result.\\

\noindent \textbf{Actions.} The number of possible actions corresponds directly to the number of available detectors in the environment.
For an environment consisting of $K$ detectors, the number of actions will be $K+2$: one action for the activation of each detector and two additional actions called ``malicious'' and ``benign.'' 
The two latter actions produce classification decisions for the analyzed file, while also terminating the analysis process. \\

\noindent \textbf{Rewards.} The rewards need to be designed so that they reflect the organizational priorities with respect to security and availability, namely in terms of the tolerance for errors in the detection process and the cost of computing resources. We use the following factors to create the reward function:

\begin{itemize}
	\item \textbf{Detection errors.} We consider two types of errors: \textit{false positive} (FP), in which a benign file is flagged as malicious (i.e., false alarm), and \textit{false negative} (FN), in which a malicious file is flagged as benign. 
	In addition to the negative rewards incurred by misclassification, it is also possible to provide a positive reward for correct classifications. 
	We elaborate on this further in Section~\ref{sec:eval}.
	
	\item \textbf{Computing resources.} In this study we chose the time required to run a detector as the approximated cost of its activation. 
	In addition to providing a close approximation of the cost of other types of resources (e.g., CPU, memory), the runtime is a clear indicator of an organization's ability to efficiently process large volumes of incoming files. 
\end{itemize}

\noindent When designing the reward function for the analysis runtime, we needed to address the large difference in this measure among various detectors. 
As shown in Table~\ref{tab:accuracy} in Section~\ref{subsec:performance}, the average runtimes can vary by orders of magnitude (from 0.7 to 44.29 seconds, depending on the detector). 
When analyzing the behavior of our model, we identified a need to mitigate these differences and encourage the use of the more computationally expensive (but also better performing) detectors.
We define the following cost function, which places an upper bound $u$ on the computational cost -- represented in our experiments by the time $t$ -- while also creating a ``smoother'' representation of the values:

\begin{equation}
C(t, u) =
\begin{cases*}
\,t & if $0\leq t\leq 1$ \\
\,min\{1 + log_2(t), u\} & if $t > 1$
\end{cases*}
\label{eq:cost}
\end{equation}

Note that while we only consider runtime as a function of the cost, our approach can be easily adapted to include additional resources such as memory usage, CPU runtime, cloud computing costs, and even electricity. 
As such, \MethodName enables organizations to \textit{easily integrate all relevant costs into their decision-making process}.

\section{EXPERIMENTAL SETUP}

\subsection{The Datasets}
\label{subsec:datasets}
We conduct our evaluation on two datasets, each consisting of a different type of executable files. 

\textbf{PE executables.} This dataset consists of 24,738 PE files, equally divided between malicious and benign. 
While we were unable to determine the creation time of each file, all files were collected from the repositories of the network security department of a large organization during October 2018; all files were obtained by the organization 1-6 months prior to our creation of the dataset. 
We used VirusTotal~\cite{total2012virustotal} to obtain our classification ground truth for training~\cite{p25,p60}. 
It is important to note that all files were known to VirusTotal prior to our experiments.

\textbf{APK executables.} This dataset consists of 20,000 APK files obtained from VirusTotal.
This dataset is also equally divided between malicious and benign, with the malicious label assigned to files flagged as such by five or more VirusTotal detectors.

\subsection{The Detectors}
\label{subsec:detectors}
Our selection of detectors was guided by three objectives:

\begin{itemize}
    \item \textbf{Off-the-shelf software.} The ability to use a malware detection solution without any special adaptation demonstrates that our approach is generic and easily applicable.
    \item \textbf{Proven detection capabilities.} By using detectors that are also in use in real-world organizations, we ensure the validity of our experiments. 
    As an example, we later describe the applications of one the detectors chosen -- Manalyze.
    \item \textbf{Runtime variance.} Since the goal of our experiments is to demonstrate \MethodName's ability to perform cost-effective detection (with runtime as our cost metric), the use of detection solutions that vary in their resource requirements was deemed preferable. 
    Moreover, such variance is consistent with real-world detection pipelines that combine multiple detector families~\cite{p102}.
\end{itemize}

\noindent Therefore, for each of our two datasets, we selected four detectors based on the objectives described above.\\

\noindent\textbf{The PE detectors:}

\textbf{pefile.} This detector uses seven features extracted from the PE header: DebugSize, ImageVersion, IatRVA, ExportSize, ResourceSize, VirtualSize2, and NumberOfSections, all presented in~\cite{p33}. 
Using those features, we trained a Decision Tree classifier.

\textbf{byte3g.} This detector uses features extracted from the raw binaries of the PE file~\cite{p31}. 
First, it constructs trigrams of bytes. 
Second, it computes the trigrams' term frequencies (TFs), which are the raw counts of each trigram in the entire file.
Third, we calculate the document frequencies (DFs), which represent the ubiquity of a trigram in the entire dataset. 
Finally, since the number of features can be substantial (up to $256^3$), we use the top 300 DF-valued features for classification. 
Using the selected features, we trained a Random Forest classifier with 100 trees.

\textbf{opcode2g.} This detector uses features based on the disassembly of the PE file~\cite{p34}. 
First, it disassembles the file and extracts the opcode of each instruction.
Second, it generates bigram representations of the opcodes. 
Third, both the TF and DF values are computed for each bigram. 
Finally, as done for byte3g, we select the 300 bigrams with the highest DF values and use their TF values as features. 
Using the selected features, we trained a Random Forest classifier with 100 trees.

\textbf{manalyze.} This detector is based on Manalyze,\footnote{https://github.com/JusticeRage/Manalyze} an open source heuristic scanning tool used by security products and companies, such as DFN-CERT\footnote{https://www.dfn-cert.de/en.html} and ANY.RUN.\footnote{https://any.run}
This detector offers multiple types of static analysis capabilities for PE files, namely plug-ins, which include: packed executable detection, ClamAV and YARA signatures, detection of suspicious import combinations, detection of cryptographic algorithms, and the verification of Authenticode signatures. 
Each plug-in returns one of three values: \textit{benign}, \textit{possibly malicious}, and \textit{malicious}. 
Since Manalyze does not offer an out-of-the-box method for combining the plugin scores, we trained a Decision Tree classifier with the scores of the plug-ins as features. \newline

\noindent\textbf{The APK detectors:}

\textbf{manifest.} This detector, suggested by Sato \etal~\cite{sato2013detecting}, detects malicious APK by analyzing its \textit{AndroidManifest.xml} file.
The authors selected six features, chosen by a heuristic scoring mechanism: permission, intent filter (action), intent filter (category), intent filter (priority), process name, and number of redefined permissions.
The textual features were converted to numeric values using a malignancy score computed based on statistical analysis.
The abovementioned features were then used to train a Decision Tree classifier.

\textbf{mmda.} Wang \etal~\cite{wang2016mmda} utilized 122 features extracted from the \textit{AndroidManifest.xml} file: the 40 most frequent permissions, 40 most frequent hardware features, 40 most frequent actions, and two additional features --- the number of permissions and the number of receiver actions. These features were used to train a Random Forest classifier with 100 trees.

\textbf{bytecode.} This detector performs frequency analysis of the Dalvik bytecode instructions~\cite{kang2013android}. 
Android apps are developed in Java, compiled to Dalvik bytecode, and executed in either Dalvik or Android Runtime virtual machines, which translate the bytecode into machine operations.
Bytecode analysis is used to analyze the app's behavior.
This detector evaluates the frequencies of the 218 bytecode instructions and uses them as features. 
Then, a Random Forest classifier with 100 trees was trained to classify the file.

\textbf{dalvikapi.} This detector uses permissions and Dalvik API calls to detect malicious APKs~\cite{chan2014static}. 
The permissions are extracted from the Android manifest file, while the API calls are extracted using the following process:
First, the \textit{classes.dex} file inside the APK is converted to a Java Archive (JAR) file. 
Second, all class files are extracted from the decompressed JAR. 
Third, the class files are decompiled into Java files that contain all API calls.
Information Gain (IG) is then applied to select the most useful permissions and API calls, keeping only those whose IG values are positive.
The authors trained several classification models using the selected features, with Random Forest achieving the best performance.
We therefore trained a Random Forest model with 500 trees.

\begin{figure*}[ht]
    \centering
    \begin{subfigure}[b]{1\textwidth}
        \centering
        \includegraphics[width=0.94\linewidth]{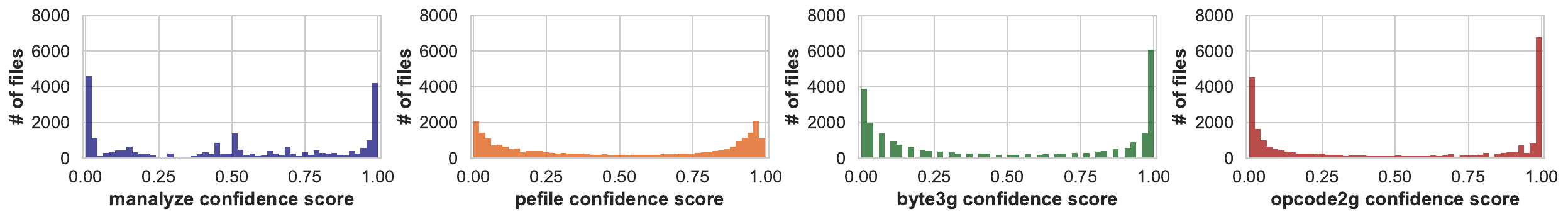}
        \vspace*{-7pt}
        \caption{PE detectors}
        \vspace*{7pt}
    \end{subfigure}%
    \newline
    \begin{subfigure}[b]{1\textwidth}
        \centering
        \includegraphics[width=0.94\linewidth]{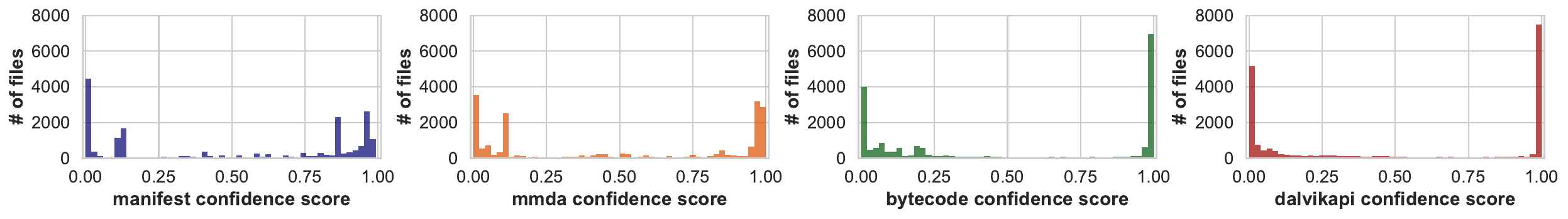}
        \vspace*{-7pt}
        \caption{APK detectors}
    \end{subfigure}
    \caption{The distributions of the files in our datasets based on the confidence score assigned to them by each detector.}
	\label{fig:toolsdist}
	\vspace*{-3pt}
\end{figure*}

\subsection{Detector Performance Analysis}
\label{subsec:performance}

We now analyze and compare the performance of the various detectors. 
This analysis has several goals: First, we aim to prove that our selected detectors are indeed effective on the datasets and that their classifications are diverse enough to support an ensemble solution. 
Secondly, we explore the effectiveness of various detector combinations and explain why an intelligent and dynamic selection of detector subsets is likely to produce near optimal performance at a much lower computational cost. 
Finally, we show that no classifier is ``dominated'' by another (see below
), thus ensuring that every detector offers the best cost-effective option in some scenarios.

In order to ensure the accuracy of the analysis presented below, we ran each detector in an isolated computer process on a dedicated machine. 
In addition, all machines were identical in their hardware and firmware configurations. \newline

\noindent \textbf{Overall detector performance.} We begin by analyzing the upper bound on the detection capability of the eight detectors (four per dataset).
All detectors were trained and tested using 10-fold cross-validation with the F1 score as the evaluation metric, and we present an average of the results. 
We define incorrect classification as a confidence threshold above 0.5 for a benign file or one that is below or equal to 0.5 for a malicious file.

For the PE detectors, the results of our analysis show that 73\% of files are classified correctly by all detectors, while only 0.65\% are mis-classified by all detectors. 
The same behavior occurs for APK files, with 78.6\% of files classified correctly by all detectors and 2\% not misclassified by all.

We derive two conclusions from these results: 
\begin{enumerate*}[a)]
    \item the number of cases where we have to use all four detectors to obtain at least a single correct classification is very small; and
    \item approximately 26.3\% of the files in the PE dataset and 19.4\% of the files in the APK dataset potentially require the use of multiple detectors to obtain correct classification.
\end{enumerate*}
We argue that these two conclusions support our rationale for using \MethodName, and that an intelligent and dynamic selection of a subset of detectors will have little effect on performance, while reducing the needed time and resources.

\begin{table}[ht]
    \vspace*{7pt}
    \centering
	\caption{The performance of the PE and APK detectors. 
	The mean running time of each detector is calculated for all files in the dataset. 
	The running times were measured on machines utilizing the same hardware (see Section~\ref{subsec:envspecs}).} 
	\label{tab:accuracy}
	\small
	\renewcommand{\arraystretch}{0.93}
	\setlength\tabcolsep{7.3pt}
	\begin{tabular}{c|c|ccHcc}
		\toprule
		\multicolumn{2}{c|}{} & F1 score (\%) & Time (sec.) & Acc. (\%) & TPR & FPR \\
		\midrule
		\multirow{4}{*}{PE} 
		& manalyze  & 83.11 &  0.75 & 82.88 & 0.844 & 0.186 \\
		& pefile    & 90.55 &  0.70 & 90.59 & 0.902 & 0.090 \\
		& byte3g    & 94.82 &  3.99 & 94.89 & 0.937 & 0.039 \\
		& opcode2g  & 95.48 & 42.99 & 95.50 & 0.951 & 0.041 \\
		\midrule
    	\multirow{4}{*}{APK} 
		& manifest  & 86.35 &  0.34 & 86.03 & 0.883 & 0.162 \\
		& mmda      & 89.76 &  0.39 & 89.82 & 0.892 & 0.096 \\
		& bytecode  & 91.75 & 10.13 & 91.99 & 0.890 & 0.051 \\
		& dalvikapi & 94.13 & 29.73 & 94.24 & 0.923 & 0.038 \\
		\bottomrule
	\end{tabular}
\end{table}

\begin{table}[ht]
    \centering
    \caption{The performance of each detector on files that were misclassified by another detector, for both datasets. 
    The results clearly show that no detector is dominated by another.}
    \vspace*{-3pt}
    \label{tab:compensate}
    \small
    \renewcommand{\arraystretch}{0.93}
    \begin{subtable}{0.47\textwidth}
        \centering
    	\caption{PE dataset}
    	\setlength\tabcolsep{10.1pt}
    	\begin{tabular}{c|cccc}
    		\toprule
    		         & manalyze & pefile  & byte3g  & opcode2g\\
    		\midrule
    		manalyze & -        & 82.96\% & 90.09\% & 91.01\% \\
    		pefile   & 68.96\%  & -       & 73.43\% & 78.93\% \\
    		byte3g   & 66.24\%  & 50.32\% & -       & 60.69\% \\
    		opcode2g & 65.71\%  & 55.90\% & 55.99\% & -       \\
    		\bottomrule
    	\end{tabular}
    \end{subtable}%
    \bigskip 
    \vspace*{2px}
    \begin{subtable}{0.47\textwidth}
        \centering
    	\caption{APK dataset}
    	\setlength\tabcolsep{9.9pt}
    	\begin{tabular}{c|cccc}
    		\toprule
    		          & manifest &  mmda   & bytecode & dalvikapi \\
    		\midrule
    		manifest  & -        & 48.43\% & 74.72\%  & 75.69\%   \\
    		mmda      & 29.39\%  & -       & 62.88\%  & 64.21\%   \\
    		bytecode  & 55.93\%  & 52.74\% & -        & 55.74\%   \\
    		dalvikapi & 41.07\%  & 36.63\% & 38.44\%  & -         \\
    		\bottomrule
    	\end{tabular}
    \end{subtable}
    \vspace*{-6pt}
\end{table}

\noindent \textbf{Absolute and relative detector performance.} Our goal in this analysis is first to present the performance (i.e., detection rate) of each detector and then to determine whether any classifier is dominated by another (thus making it redundant, unless it is more computationally efficient).
We begin our analysis by presenting the absolute performance for our two datasets.
Table~\ref{tab:accuracy} presents the performance of the detectors for both datasets with regard to the mean F1-score, true positive (malware detection) rate, and false positive (misclassification of benign files) rate. 
The F1-scores of the PE detectors ranges from 83.11 to 95.48\%, and that of the APK detectors ranges from 86.35 to 94.13\%. 
For both datasets, the more computationally-expensive detectors generally perform better.

Next, we attempt to determine whether any detector is dominated by another (i.e., if one detector produces classifications that are equal to or better than those of another detector for any given file).
Table~\ref{tab:compensate} shows the results of this analysis for the PE and APK detectors. 
The value of each cell indicates what percentage of the files that were \textit{incorrectly classified} by one detector (row) where \textit{correctly classified} by another (column).
For example, the pefile detector was able to correctly classify 50.32\% of the errors of the byte3g detector.
Therefore, as seen in the tables, \textit{no detector is being dominated by another}.
Moreover, the large variance in the detection rates of detectors for misclassified files shown in Table~\ref{tab:compensate} further suggests that an intelligent selection of detector subsets --- where detectors complement each other --- can yield a high F1 score. Our conclusions are further supported by our analysis in Appendix \ref{apndx:compensate_fp}, which repeats the experiments using a different decision threshold. \newline

\begin{table*}[htbp!]
    \footnotesize
    \centering
	\caption{A summary of the performance of the detector combinations, showing only the aggregation methods that obtained the highest F1 scores for each combination.
	The full results are presented in Appendix~\ref{apndx:full_baselines}  (Tables~\ref{tab:combinations_pe_full}).
	The results are calculated using 10-fold cross-validation and sorted in a descending order according to their mean F1 score. 
	We also presented the results of our experiments, described in Section~\ref{subsec:experimental_results}.
	Aggregations were conducted using: \textit{majority}, \textit{or}, \textit{stacking} using \textit{Decision Tree and Random Forest} methods (see Section~\ref{subsec:performance}).
    }
	\label{tab:combinations_all}
	\renewcommand{\arraystretch}{0.93}
	\setlength\tabcolsep{6.5pt}
	\begin{tabular}{cclcccccccc}
		\toprule
		Type & \# & Detector Combination & Aggregation & F1 score (\%) & Time (sec) & Precision (\%) & Recall (\%) & Accuracy (\%) & FP (\%) & FN (\%) \\
		\midrule
		\rowcolor{olive!20}
		\cellcolor[gray]{1} \multirow{20}{*}{PE} & - &
        Experiment \#1                   &     SPIREL     & 96.865 &  48.61 & 97.23 & 96.51 & 96.867 &  1.38 &  1.75 \\
        & (\ref{bl_pe:1}) &
        manalyze,pefile,byte3g,opcode2g  & stacking (RF)  & 96.852 &  49.73 & 96.94 & 96.76 & 96.859 &  1.52 &  1.62 \\
        \rowcolor{lightgray!40} \cellcolor[gray]{1} & - &
        Experiment \#2                   &     SPIREL     & 96.812 &  48.37 & 96.45 & 97.18 & 96.801 &  1.79 &  1.41 \\
        & (\ref{bl_pe:2}) &
        manalyze,byte3g,opcode2g         &    majority    & 96.693 &  49.03 & 97.07 & 96.32 & 96.709 &  1.45 &  1.84 \\
        & (\ref{bl_pe:5}) &
        byte3g,opcode2g                  &    majority    & 96.358 &  48.28 & 96.68 & 96.04 & 96.374 &  1.65 &  1.98 \\
        & (\ref{bl_pe:6}) &
        pefile,byte3g,opcode2g           &    majority    & 96.279 &  48.98 & 96.74 & 95.82 & 96.301 &  1.61 &  2.09 \\
        \rowcolor{olive!20} \cellcolor[gray]{1} & - &
        Experiment \#3                   &     SPIREL     & 96.225 &  10.53 & 96.09 & 96.36 & 96.212 &  1.96 &  1.82 \\
        & (\ref{bl_pe:7}) &
        manalyze,pefile,opcode2g         & stacking (RF)  & 96.202 &  45.74 & 96.66 & 95.75 & 96.224 &  1.65 &  2.12 \\
        & (\ref{bl_pe:11}) &
        manalyze,pefile,byte3g           &    majority    & 95.591 &   5.44 & 96.06 & 95.13 & 95.618 &  1.95 &  2.43 \\
        & (\ref{bl_pe:13}) &
        manalyze,opcode2g                &    majority    & 95.543 &  45.04 & 95.73 & 95.35 & 95.557 &  2.12 &  2.32 \\
        & (\ref{bl_pe:16}) &
        opcode2g                         &      none      & 95.475 &  44.29 & 95.83 & 95.13 & 95.497 &  2.07 &  2.43 \\
        & (\ref{bl_pe:18}) &
        pefile,opcode2g                  &    majority    & 95.419 &  44.99 & 95.84 & 95.01 & 95.444 &  2.06 &  2.49 \\
        \rowcolor{lightgray!40} \cellcolor[gray]{1} & - &
        Experiment \#4                   &     SPIREL     & 95.315 &   3.68 & 97.77 & 92.98 & 95.424 &  1.06 &  3.51 \\
        & (\ref{bl_pe:21}) &
        manalyze,byte3g                  &    majority    & 95.143 &   4.74 & 95.14 & 95.14 & 95.149 &  2.43 &  2.43 \\
        & (\ref{bl_pe:25}) &
        byte3g                           &      none      & 94.823 &   3.99 & 95.98 & 93.69 & 94.890 &  1.96 &  3.15 \\
        & (\ref{bl_pe:26}) &
        pefile,byte3g                    &    majority    & 94.813 &   4.69 & 95.30 & 94.33 & 94.846 &  2.32 &  2.83 \\
        & (\ref{bl_pe:37}) &
        manalyze,pefile                  &    majority    & 92.336 &   1.45 & 92.96 & 91.72 & 92.396 &  3.47 &  4.14 \\
        \rowcolor{lightgray!40} \cellcolor[gray]{1} & - &
        Experiment \#5                   &     SPIREL     & 91.065 &   0.73 & 92.71 & 89.48 & 91.220 &  3.52 &  5.26 \\
        & (\ref{bl_pe:39}) &
        pefile                           &      none      & 90.552 &   0.70 & 90.88 & 90.23 & 90.597 &  4.52 &  4.88 \\
        & (\ref{bl_pe:48}) &
        manalyze                         &      none      & 83.113 &   0.75 & 81.89 & 84.38 & 82.876 &  9.32 &  7.80 \\
        \midrule
        
        \rowcolor{olive!20} 
        \cellcolor[gray]{1} \multirow{20}{*}{APK} & - &
        Experiment \#1                   &     SPIREL     & 94.369 &  31.49 & 95.07 & 93.68 & 94.409 &  2.43 &  3.16 \\
        & (\ref{bl_pe:1}) &
        manifest,mmda,bytecode,dalvikapi & stacking (RF)  & 94.303 &  40.59 & 96.04 & 92.63 & 94.399 &  1.91 &  3.69 \\
        \rowcolor{lightgray!40} \cellcolor[gray]{1} & - &
        Experiment \#2                   &     SPIREL     & 94.208 &  30.69 & 96.64 & 91.90 & 94.353 &  1.60 &  4.05 \\
        & (\ref{bl_pe:2}) &
        bytecode,dalvikapi               &    majority    & 94.194 &  39.86 & 96.53 & 91.97 & 94.326 &  1.66 &  4.02 \\
        & (\ref{bl_pe:3}) &
        dalvikapi                        &      none      & 94.131 &  29.73 & 96.07 & 92.27 & 94.242 &  1.89 &  3.87 \\
        & (\ref{bl_pe:4}) &
        manifest,bytecode,dalvikapi      & stacking (RF)  & 94.013 &  40.20 & 96.06 & 92.05 & 94.133 &  1.89 &  3.98 \\
        & (\ref{bl_pe:5}) &
        mmda,bytecode,dalvikapi          & stacking (RF)  & 93.975 &  40.25 & 95.76 & 92.25 & 94.080 &  2.04 &  3.88 \\
        \rowcolor{olive!20} \cellcolor[gray]{1} & - &
        Experiment \#3                   &     SPIREL     & 93.919 &   8.73 & 94.24 & 93.60 & 93.971 &  2.86 &  3.20 \\
        & (\ref{bl_pe:8}) &
        manifest,mmda,dalvikapi          & stacking (RF)  & 93.760 &  30.46 & 95.59 & 92.00 & 93.871 &  2.13 &  4.00 \\
        & (\ref{bl_pe:11}) &
        mmda,dalvikapi                   &    majority    & 93.395 &  30.12 & 94.70 & 92.13 & 93.480 &  2.58 &  3.94 \\
        & (\ref{bl_pe:13}) &
        manifest,dalvikapi               & stacking (RF)  & 93.253 &  30.07 & 94.86 & 91.70 & 93.359 &  2.49 &  4.15 \\
        & (\ref{bl_pe:21}) &
        manifest,bytecode                &    majority    & 92.335 &  10.47 & 93.38 & 91.31 & 92.414 &  3.24 &  4.35 \\
        & (\ref{bl_pe:23}) &
        mmda,bytecode                    &    majority    & 92.286 &  10.52 & 94.21 & 90.44 & 92.435 &  2.78 &  4.79 \\
        & (\ref{bl_pe:26}) &
        manifest,mmda,bytecode           & stacking (RF)  & 92.140 &  10.86 & 94.30 & 90.08 & 92.309 &  2.73 &  4.96 \\
        \rowcolor{lightgray!40} \cellcolor[gray]{1} & - &
        Experiment \#4                   &     SPIREL     & 92.021 &   3.29 & 91.78 & 92.26 & 92.001 &  4.13 &  3.87 \\
        & (\ref{bl_pe:28}) &
        bytecode                         &      none      & 91.754 &  10.13 & 94.63 & 89.05 & 91.991 &  2.53 &  5.48 \\
        \rowcolor{lightgray!40} \cellcolor[gray]{1} & - &
        Experiment \#5                   &     SPIREL     & 91.113 &   1.63 & 89.60 & 92.68 & 90.956 &  5.38 &  3.66 \\
        & (\ref{bl_pe:37}) &
        manifest,mmda                    & stacking (RF)  & 90.035 &   0.73 & 90.70 & 89.38 & 90.099 &  4.59 &  5.31 \\
        & (\ref{bl_pe:41}) &
        mmda                             &      none      & 89.764 &   0.39 & 90.31 & 89.23 & 89.817 &  4.79 &  5.39 \\
        & (\ref{bl_pe:48}) &
        manifest                         &      none      & 86.346 &   0.34 & 84.50 & 88.28 & 86.029 &  8.10 &  5.87 \\
        \bottomrule
	\end{tabular}
	\vspace*{-5pt}
\end{table*}

\noindent \textbf{Confidence score distributions of detectors.} Next, we analyze the detectors' confidence score distributions. 
Our goal in this analysis is to determine whether the detectors are capable of nuanced analysis; we hypothesize that detectors which produce multiple values on the [0,1] scale (rather than only zeros and ones) are more likely to enable our DRL approach to devise more nuanced strategies for selecting detector combinations.

The results of our analysis are presented in Figure~\ref{fig:toolsdist}. 
While it is clear that all detectors assign either zeros or ones to the majority of files, a large number of files (particularly for less expensive detectors with lower F1 scores) are assigned intermediary values. 
We therefore conclude that the classifications produced by the detectors are sufficiently diverse to support a nuanced DRL policy. 
The efficacy of our approach is evaluated by our experiments in Section~\ref{subsec:experiments}. \newline

\noindent \textbf{Performance and time usage of detector combinations.} 
Finally, we provide a comprehensive analysis of the performance and time consumption for all possible detector combinations. A summary of the results is presented in Table~\ref{tab:combinations_all}, and the full results are presented in Tables~\ref{tab:combinations_pe_full}~and~\ref{tab:combinations_apk_full} in Appendix~\ref{apndx:full_baselines}.
To evaluate the performance of each combination, we aggregated the confidence scores using three different methods presented in~\cite{p95}:

\begin{itemize}
    \item \textbf{or} -- classifies a file as malicious if any of the participating detectors classifies it as such (yields a score of 0.5 and above). 
    This method improves detection sensitivity, but at the cost of a higher false positive percentage. 
    \item \textbf{majority} -- uses voting to classify the files. 
    \item \textbf{stacking} -- a ML model is used to produce the final classification. Individual confidence scores are used as input features. 
    In our evaluation, we used two types of classifiers -- Decision Tree (DT) and Random Forest (RF) -- and evaluated each using 10-fold cross-validation.
\end{itemize}

\noindent Interestingly, our analysis of PE detectors shows that in the case of \textit{majority}, the optimal performance is not achieved by combining all classifiers but rather by combining just three of them.
Furthermore, some detector combinations (manalyze, pefile, byte3g) for PE files and (bytecode,dalvikapi) for APK files outperform other detector combinations while also being more computationally efficient. 
The results further support our claim that an intelligent selection of detector combinations is highly beneficial.

\subsection{Evaluation Environment}
\label{subsec:envspecs}

\begin{figure}[htp!]
  \centering
  \includegraphics[scale=0.31]{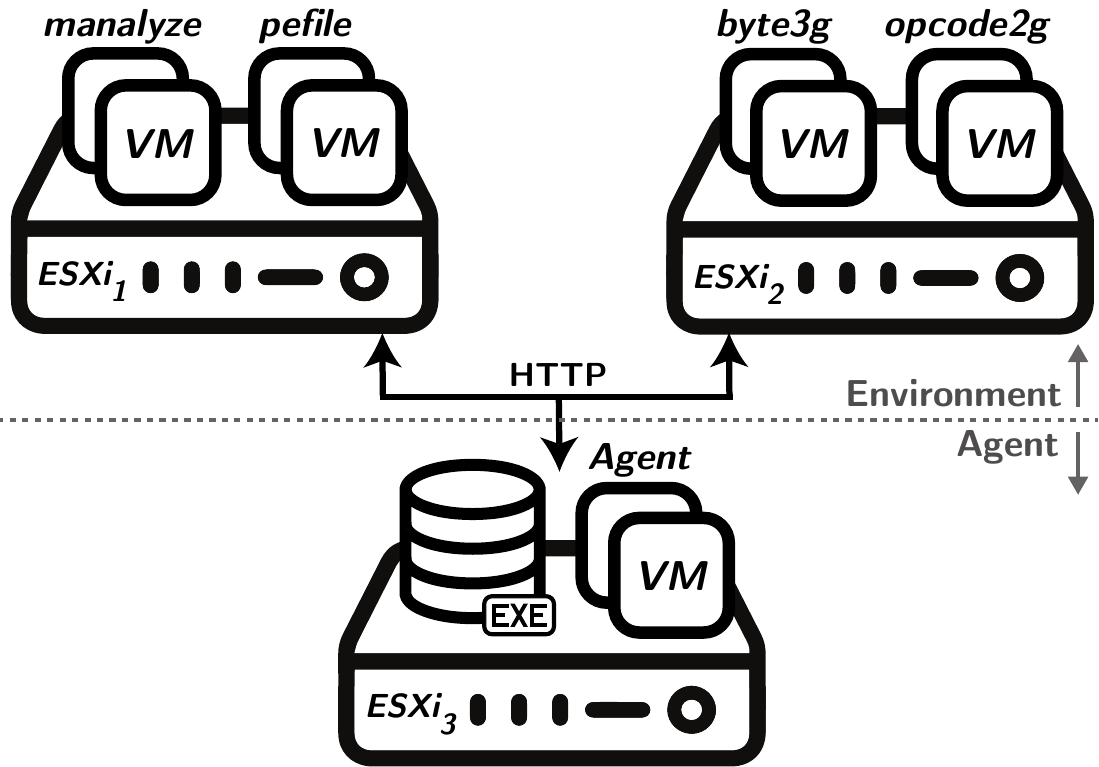}
  \caption{The experimental infrastructure for the PE experiments (the same was used for the APK experiments, using different detectors and agents). $ESXi_1$ hosts manalyze and pefile detectors, $ESXi_2$ hosts byte3g and opcode2g detectors, and $ESXi_3$ hosts the agent and the PE corpus.} 
  \label{fig:infrastructure}
\end{figure}

Figure~\ref{fig:infrastructure} illustrates the environment used for the PE dataset. 
The same infrastructure is used for the APK dataset, using different detectors. 
For every experiment we used three VMware ESXi servers, each containing two processing units (CPUs), with 32 cores, 512GB of RAM, and 100TB of SSD disk space. 
Two servers were used to host the environment and its detectors, while the remaining server hosted our DRL agent.
We deployed two detectors on each server. 
This deployment setting can easily be extended to include additional detectors or replicate the existing ones to increase the throughput. 
Our main goal in setting up the environment was to demonstrate a large-scale implementation which is both scalable and flexible, thus ensuring its relevance to real-world scenarios. 

Both the agent processes and the detectors run on virtual machines with the Ubuntu 18.04 LTS operating system.
Each virtual machine has four CPU cores, 16GB of RAM, and 100GB of SSD storage.
The agent uses a management service that allows both the training and execution of the DRL algorithm, using different tuning parameters. 
Upon the arrival of a file for analysis, the agent stores it in a dedicated storage space, which is also accessible to all detectors running in the environment. 
The agent also uses external storage to store file and detector-based features, all logging information, and the analysis output. 
All of this information is later indexed by an analytics engine.

Training times of both the APK and PE agents were relatively short, ranging between 181-285 minutes. A comprehensive description of the training time analysis is presented in Appendix~\ref{apndx:training_times}. The number of epochs needed for the models to converge ranged from four to six epochs. The main reason for the short training times is small size of the input. It is worth noting that the addition of more detectors (i.e., more actions, inputs and outputs) will only cause a linear increase in the size of the network. For this reason, our can easily scale to larger detection environments.

\subsection{Experimental Setting}

The following settings were used in all of our experiments:
\begin{itemize}[leftmargin=*]
    \item We used 10-fold cross-validation, with label ratios maintained.
    \item We implemented the framework using Python v3.6, with the ChainerRL~\cite{nandy2018reinforcement} deep reinforcement learning library to create and train the agent. The environment was implemented using the OpenAI Gym~\cite{gym}.
    \item We used the actor-critic~\cite{p96} with experience replay~\cite{p97} (ACER) RL algorithm to produce our security policies. 
    We set the size of the replay buffer to 5,000, and begin using it in the training process after 10,000 episodes.
    \item Both the policy and action-value networks of the ACER consist of the following architecture: input layer of size four (the state vector's size), a hidden layer of size 20, and an output layer of size six (equal to the number of possible actions: four detectors and two final classifications). 
    All layers except for the output layer use the ReLU activation function, while the latter uses softmax. 
    \item We set our initial learning rate to $7e-4$, with an exponential decay rate of $0.99$ and a fuzz factor (epsilon) of $1e-2$. 
    Our chosen optimizer was RMSprop~\cite{rmsprop}. 
    In all experiments, our models were trained until convergence.
    \item In order to discourage the agent from querying the same detector twice (an obvious waste of resources, since no new information is gained), we specify that such actions incur a very large cost of -10,000. 
    The same penalty applies to attempts to classify a file without using even a single detector.
    \item We used the cost function presented in Equation~\ref{eq:cost}. For the PE dataset the upper bound parameter $u$ was six -- due to the computation cost of the data, which was less than 34 seconds for 95\% of the data and resulted in $1+log_2(34)=6.09$. 
    For the APK dataset the upper bound parameter $u$ was five -- due to the computation cost of the data, which was less than 28 seconds for 95\% of the data and resulted in $1+log_2(28)=4.81$.
    \item As in other recent studies in the field of malware detection~\cite{p35, deepmalnet, metagraph2vec, abderrahmane2019android}, we used precision/recall as our evaluation measure. However, since \MethodName applies cost-effective analysis, we also need to take into account the \textit{element of time}. We therefore use the F1 score, which combines precision and recall into a single measure and enables us to present the F1/time trade-offs of both \MethodName and the baselines. The F1 score is calculated as follows:
    \begin{equation}
        F1 = 2 \cdot \frac{Precision \cdot Recall}{Precision + Recall}
    \end{equation}
  
\end{itemize}

\section{EVALUATION}
\label{sec:eval}

Our evaluation is designed to assess the efficacy and efficiency of the proposed approach in a range of real-world scenarios. 
We begin by presenting the reward functions (i.e., organizational policies with various performance/resources trade-off preferences) designed for our experiments. 
We then evaluate \MethodName on the two datasets described in Section \ref{subsec:datasets}. 
Our results and analysis in Sections \ref{subsec:experimental_results} and \ref{subsec:analysis} show that regardless of the desired performance/efficiency trade-off, \MethodName outperforms all relevant baselines.

Next, we evaluate our approach's performance in three real-world scenarios. 
We begin by analyzing \MethodName's ability to adapt to different costs for different types of mistakes (false positives vs. false negatives). 
Our conclusions, presented in Section \ref{subsec:sensitivity}, show that our approach easily adapts its policy to accommodate different organizational preferences. 
Next, we evaluate the effectiveness of our approach in the context of transfer learning. 
This setting is particularly important in cases where little data is available to train an ML model, thus necessitating the use of models trained on other datasets and domains.
Our results in Section~\ref{subsec:transfer}, show that a model of \MethodName trained on one dataset can easily and effectively be transferred to another. 
Moreover, such a scenario usually leads to improved performance and reduced training times. 
Finally, in Section~\ref{subsec:adapting}, we explore \MethodName's ability to adapt to changes in the performance of some of its detectors (e.g., when a detector becomes ineffective because of a new obfuscation technique). 
We show that \MethodName is highly adaptable and requires very little retraining.

\begin{figure*}[ht]
    \centering
    \begin{subfigure}{.49\textwidth}
      \centering
      \includegraphics[scale=0.45]{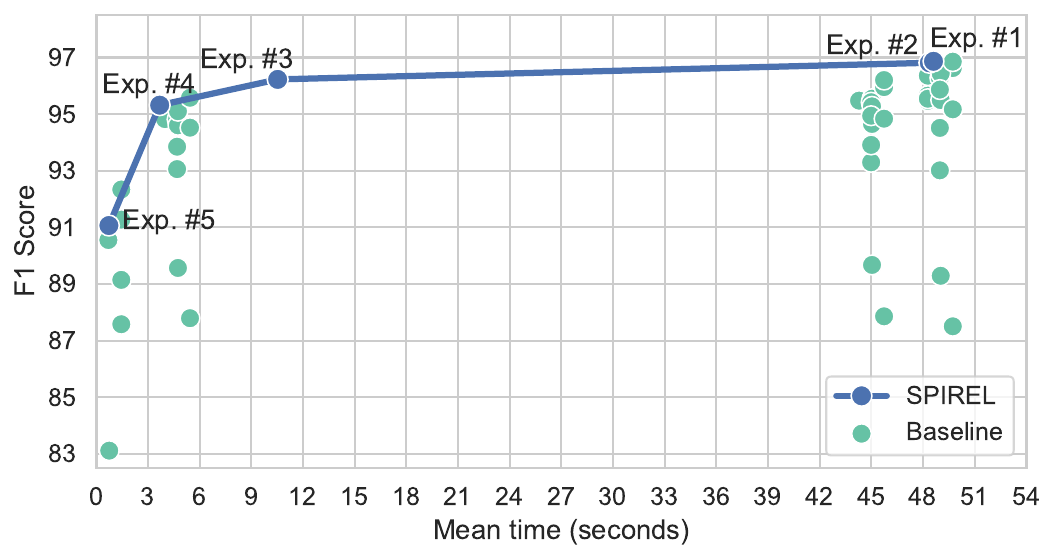}
      \vspace*{-5pt}
      \caption{PE dataset}
    \end{subfigure}
    \begin{subfigure}{.49\textwidth}
      \centering
      \includegraphics[scale=0.45]{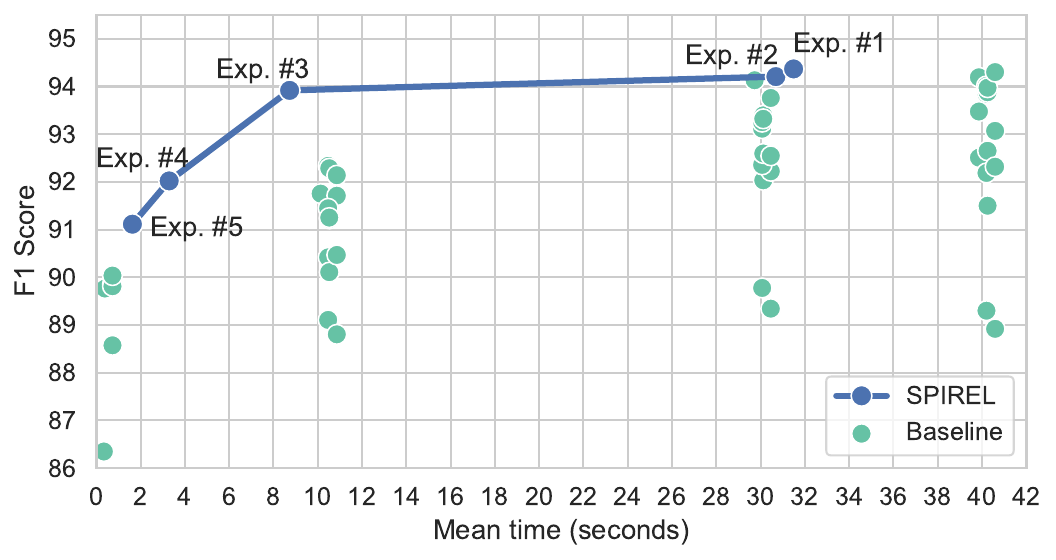}
      \vspace*{-5pt}
      \caption{APK dataset}
    \end{subfigure}
  \caption{The results of the experiments conducted on each dataset, compared to the results of their corresponding baselines, with regard to the F1 score and mean time per file.
  Our proposed solution generates a Pareto frontier for both cases.} 
  \label{fig:pareto}
\end{figure*}

\begin{table}[ht]
  \vspace*{7pt}
  \caption{The reward setup of our PE and APK experiments in relation to the costs. The functions $C'(t)=C(t, 6)$ and $C''(t)=C(t, 5)$ are based on the function $C(t, u)$ defined in Equation~\ref{eq:cost}.}
  \small
  \label{tab:experiments}
  \renewcommand{\arraystretch}{0.94}
  \setlength\tabcolsep{5.5pt}
  \begin{tabular}{c|c|cccc|cc}
    \toprule
    \multicolumn{2}{c|}{Exper.} & \multicolumn{4}{c|}{Reward Setup} & F1 score & Time\\
    \multicolumn{2}{c|}{\#} & TP & TN & FP & FN & (\%) & (sec)\\
    \midrule
    \multirow{5}{*}{PE}
    & 1 & C'(t)  & C'(t)  & -C'(t)   & -C'(t)   & 96.865 & 48.61 \\
    & 2 & C'(t)  & C'(t)  & -10C'(t) & -10C'(t) & 96.812 & 48.37 \\
    & 3 & 1      & 1      & -C'(t)   & -C'(t)   & 96.225 & 10.53 \\
    & 4 & 10     & 10     & -C'(t)   & -C'(t)   & 95.315 & 3.68  \\
    & 5 & 100    & 100    & -C'(t)   & -C'(t)   & 91.065 & 0.73  \\
    \midrule
    \multirow{5}{*}{APK}
    & 1 & C''(t) & C''(t) & -C''(t)   & -C''(t)   & 94.369 & 31.49 \\
    & 2 & C''(t) & C''(t) & -10C''(t) & -10C''(t) & 94.208 & 30.69 \\
    & 3 & 1      & 1      & -C''(t)   & -C''(t)   & 93.919 &  8.73 \\
    & 4 & 10     & 10     & -C''(t)   & -C''(t)   & 92.021 &  3.29 \\
    & 5 & 100    & 100    & -C''(t)   & -C''(t)   & 91.113 &  1.63 \\
  \bottomrule
\end{tabular}
\vspace*{-10pt}
\end{table}

\subsection{Experiments}
\label{subsec:experiments}
We define five reward functions, each placing a different emphasis on the classification/efficiency trade-offs. 
The reward functions -- each representing a different set of organizational priorities -- assign different costs to correct/incorrect file classification as well as to the computational cost involved.

The composition of each reward function is presented in Table~\ref{tab:experiments}, along with its overall F1 score and mean running time. 
It is important to note that the computational cost of using a detector is never calculated independently, but rather as a function of correct/incorrect file classification (we elaborate on this point below). 
Finally, we wish to point out that the computational costs of the detectors are calculated using the average execution time of the files we used for training. 
This practice enables the algorithm to converge faster.

\noindent For each of our two datasets, we ran five sets experiments (a set for each of our reward functions).
Next, we describe our five experiments and the rationale of their reward functions.\newline

\noindent \textbf{Experiments \#1 and \#2.} In these experiments, the reward for both correct and incorrect classification is dependent on the resources used, approximated by running time, to reach the decision. 
We explore two variants: one where the cost of incorrect classification is opposite and equal to that of correct classification (experiment \#1), and another where the cost of incorrect classification is 10 times that of the reward for correct classification (experiment \#2). 

It is important to note that experiments \#1 and \#2 are oriented more towards detection (i.e., high F1 scores) than efficiency. 
By assigning higher rewards for correct classifications \textit{that took more time to complete} we encourage our DRL agent to query every detector that is likely to increase the amount of available information on an analyzed file. 
The DRL agent is only likely to refrain from querying a detector when \textit{a)} the agent has high uncertainty regarding the classification of the analyzed file, \textit{b)} the detector is unlikely to provide any useful information. In this setting, querying additional detectors is considered risky, because it will make a misclassification more costly.

\noindent \textbf{Experiments \#3, \#4, and \#5.} Here we explore reward function configurations where the rewards assigned to correct classifications are fixed while the cost of incorrect classification depends on the amount of computing resources used to reach the final decision. 
We explore three variants of this approach, where the rewards for correct classifications vary by orders of magnitude (1, 10, and 100). 

This set of experiments is oriented more towards efficiency than experiments \#1 and \#2. 
The reason for this is the \textit{asymmetric reward structure} we apply here: since the positive reward for correct classifications is fixed, the DRL agent has no incentive to query detectors that are likely to provide little or no new information. Querying such detectors will only make the cost of making an incorrect classification more ``painful,'' because the penalties are proportional to the overall running time. 
In other words, this reward structure pushes the DRL agent towards querying the most computationally-efficient subset of detectors needed in order to obtain a correct classification.

\subsection{Experimental Results}
\label{subsec:experimental_results}

As explained in Section \ref{sec:introduction}, no current ML approach exists for the dynamic selection of detector subsets. For this reason, any baseline we compare our approach to will have to consist of a preselected detector subset (i.e., defined prior to running). We therefore compare \MethodName \textit{to all possible detector subset combinations and their aggregation methods}. Given that we have four detectors and three aggregation methods (or, majority, and stacking), we evaluated 48 baselines for each of our datasets.

The results of our experiments for both datasets are shown in Table~\ref{tab:combinations_all}, which presents the F1 score and average runtime for all runs. 
Please note that due to the large number of baselines (48 per dataset), Table~\ref{tab:combinations_all} mainly presents the top performing baselines. The complete evaluation results are presented in Appendix~\ref{apndx:full_baselines}. An analysis of the results shows that \MethodName consistently outperforms all of the baselines. More specifically, for every desired level of performance (with the exception of the poorest performing baselines which consist of a single detector), there exists a reward-function configuration for \MethodName that can offer an equal or better performance at a reduced runtime. 

Our conclusion that \MethodName outperforms all baselines regardless of the desired performance/efficiency trade-off is further supported by Figure~\ref{fig:pareto}. 
By plotting all baselines and \MethodName experiments on a 2D-graph it is clear that regardless of the baseline, there is always a reward-function configuration of our approach that offers equal or better performance at equal or shorter running time. Therefore, the results of our approach on the various experiments form the \textit{Pareto frontier} \cite{kim2006adaptive} for both evaluated datasets.

\MethodName's superior performance is particularly evident when compared to the top performing baselines. For both datasets, our approach outperformed the top performing baselines (96.865\% compared to 96.852\% and 94.369\% to 94.303\% for the PE and APK experiments, respectively) while also being more computationally efficient (48.61 compared to 49.73 seconds and 31.49 to 40.59 seconds for the PE and APK, respectively). These results were obtained in experiment \#1, whose reward function is geared towards higher F1 scores rather than efficiency. Alternatively, \MethodName was able to reduce the running time by \textasciitilde80\% (10.52 compared to 49.73 seconds and 8.73 to 40.59 seconds for the PE and APK, respectively) while reducing the F1 score by \textasciitilde0.5\% (96.225\% compared to 96.852\% and 93.919\% to 94.303\% for the PE and APK experiments, respectively). These results were obtained in experiment \#3, whose reward function is geared towards higher efficiency.

Our experiments clearly demonstrate that the organizational security policy can be effectively managed using different cost/reward combinations. 
Moreover, it is clear that the use of DRL offers much greater flexibility in shaping the security policy than simple tweaking of the confidence threshold (the only available method for most ML-based detection algorithms). To strengthen this notion, we show in Appendix \ref{apndx:pareto_acc} that \MethodName forms the Pareto Frontier using the accuracy metric as well. Finally, we wish to reemphasize that \textit{these detector combinations are not chosen in advance}. Instead, they are chosen iteratively, with the confidence score of the already-applied detectors used to guide the policy's next step.

\begin{figure}[htp]
    \centering
    \begin{subfigure}{.47\textwidth}
      \centering
      \includegraphics[scale=0.5]{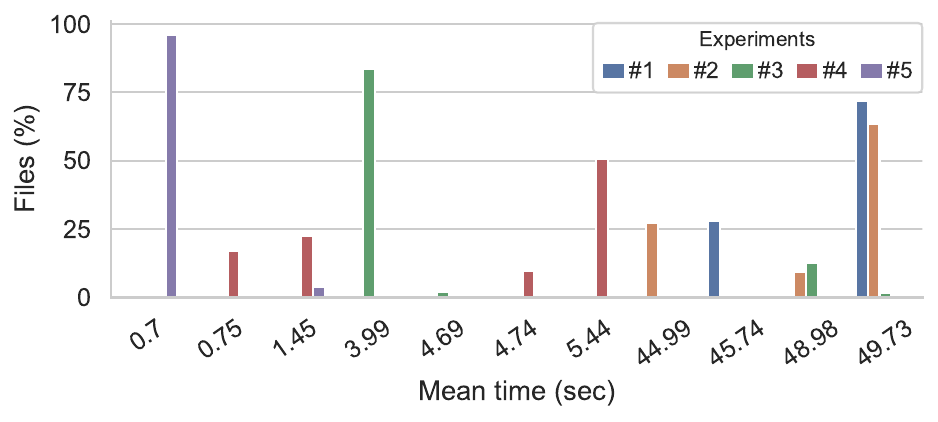}
      \caption{PE experiments}
      \vspace{1mm}%
    \end{subfigure}
    \begin{subfigure}{.47\textwidth}
      \centering
      \includegraphics[scale=0.5]{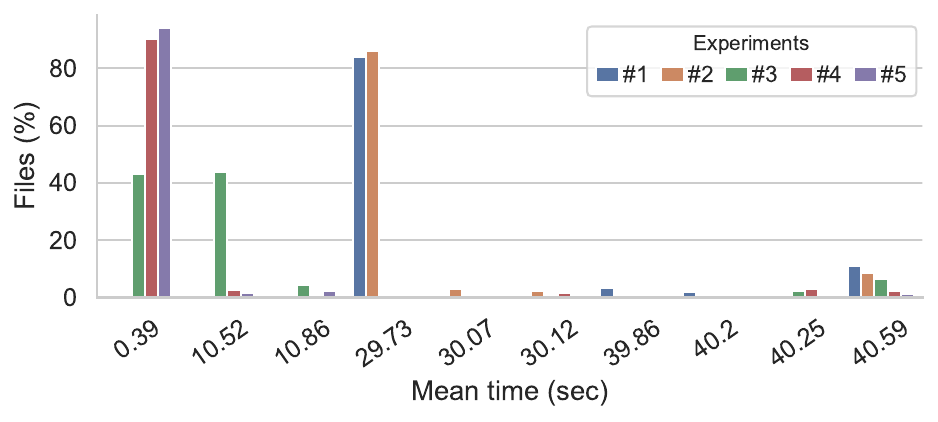}
      \caption{APK experiments}
    \end{subfigure}
  \caption{Distribution of the detector subsets (represented by their cumulative runtime) chosen by the DRL-agent in each experiment of both datasets.}
  \label{fig:expdist}
\end{figure}

\subsection{Analysis: DRL Agent Action Selection}
\label{subsec:analysis}

In the previous section we showed that the choice of the reward function, i.e., the relative priorities assigned to performance vs. efficiency, plays a significant role in \MethodName's selection of detectors. 
More specifically, the use of DRL enables our approach to craft a personalized set of detectors for each file. 
We now perform an in depth analysis of the detector selection strategies of our experiments.

Figure~\ref{fig:expdist} presents the distribution of detector combinations for each experiment (each combination is represented by its average running time on the x-axis). 
The differences between the experiments are clear: experiments \#1 and \#2 (with the performance-oriented reward functions) utilize only computationally expensive detector combinations, while the other experiments (e.g., experiment \#4) utilize a larger number of detector combinations with varying degrees of computational cost. 
This diversity helps explain \MethodName's ability to achieve a high F1 score at a significantly reduced computational cost: for what it assumes to be easy to classify cases, our approach takes a calculated risk and only queries a small number of computationally-efficient detectors.

Please note that a comprehensive breakdown of the detector subset selection of each experiment can be found in Appendix~\ref{apndx:drl_agent_action_distribution}.

\begin{table}[bp!]
  \centering
  \small
  \caption{The reward setup of our sensitivity experiments using the PE dataset. The function $C'(t)=C(t, 6)$ is based on the function $C(t, u)$ defined in Equation~\ref{eq:cost}.}
  \label{tab:sensativity_reward_setup}
  \setlength\tabcolsep{7.3pt}
  \begin{tabular}{c|c|cc|cc|cc}
    \toprule
    \multicolumn{2}{c|}{Exper.} & \multicolumn{2}{c|}{Cost Function} & F1 score & Time & FP & FN \\
    \multicolumn{2}{c|}{} & FP & FN & (\%) & (sec) & (\%) & (\%)\\
    \midrule
    \multirow{4}{*}{PE}
    & 3 & -C'(t)   & -C'(t)     & 96.225 & 10.53 & 1.96 & 1.82 \\
    \cmidrule{2-8}
    & 6 & -C'(t)   & -10C'(t)   & 96.144 & 15.43 & 2.79 & 1.13 \\
    & 7 & -C'(t)   & -50C'(t)   & 95.440 & 18.67 & 3.66 & 1.02 \\
    & 8 & -C'(t)   & -100C'(t)  & 93.919 & 25.52 & 5.47 & 0.89 \\
  \bottomrule
\end{tabular}
\end{table}

\begin{table}[bp!]
    \small
    \caption{Distribution of detector combination choices made by the agent of the PE dataset, for the experimental policies of the sensitivity analysis, as well as for experiment \#3.}
	\label{tab:expactiondist_sensativity_pe}
	\setlength\tabcolsep{3.8pt}
	\begin{tabular}{c|c|lcc}
		\toprule
		Exper. & F1 (\%) & Action Sequences & Time (sec) & Files (\%)\\
		\midrule
        3 & 96.225
          & byte3g                          &  3.99 & 83.38 \\
         && byte3g,pefile,opcode2g          & 48.98 & 12.67 \\
         && byte3g,pefile                   &  4.69 &  2.15 \\
         && byte3g,pefile,opcode2g,manalyze & 49.73 &  1.80 \\
        \midrule
        6 & 96.144
          & byte3g                          & 3.99 & 74.28  \\
         && byte3g,opcode2g                 & 48.28 & 19.88 \\
         && byte3g,opcode2g,manalyze        & 49.03 & 5.84  \\
        \midrule 
        7 & 95.440
          & manalyze,pefile,byte3g          &  5.44 & 70.14 \\
         && manalyze,pefile,byte3g,opcode   &  49.73 & 29.86 \\
        \midrule
        8 & 93.919
          & pefile,byte3g                    &  4.69 & 53.11 \\
          && pefile,byte3g,opcode2g          &  48.98 &  38.62 \\
          && pefile,byte3g,opcode2g,manalyze &  49.73 & 8.27 \\
        \bottomrule
	\end{tabular}
\end{table}

\begin{table*}[ht]
  \small
  \centering
  \caption{Transfer learning analysis based on experiment \#3 of both datasets. In addition to the convergence information and performance analysis, we analyzed the influence of various training set sizes on transfer learning.}
  \label{tab:transfer_analysis}
  \renewcommand{\arraystretch}{0.9}
  \setlength\tabcolsep{5.3pt}
  \begin{tabular}{c|cr|cc|ccccccc}
   \toprule
    Dataset & \multicolumn{2}{c|}{Experiment} & \multicolumn{2}{c|}{Training} & \multicolumn{7}{c}{Performance}\\
    Size (\%) & Type   & \multicolumn{1}{c|}{Name} & Epochs (\#)  & Time (min.)  & F1 score (\%) & Time (sec.) & Precision (\%) & Recall (\%) & Accuracy (\%) & FP (\%) & FN (\%) \\
    
    \midrule
    \rowcolor{lightgray!40}
    \cellcolor[gray]{1} 
    & \#3  & PE                 & 5 & 242.96      & \textbf{96.225}   & 10.53       & 96.09     & \textbf{96.36}  &\textbf{ 96.212}        & 1.96    & \textbf{1.82}    \\
    \rowcolor{lightgray!40}
    \cellcolor[gray]{1} 
    & TL   & APK$\rightarrow$PE & \textbf{3} & \textbf{133.72}     & 96.182   & \textbf{10.46}       & \textbf{96.12}     & 96.24  & 96.178        & \textbf{1.94 }   & 1.88    \\
    & \#3  & APK                  & 4            & 187.45      & 93.919   & \textbf{8.73}        & 94.24     & 93.60  & 93.971        & 2.86    & 3.20    \\
    \multirow{-4}{*}{100} 
    & TL   & PE$\rightarrow$APK & \textbf{2}    & \textbf{98.13}       & \textbf{94.067}   & 9.27        & \textbf{94.27}     & \textbf{93.86}  & \textbf{94.082}        & \textbf{2.85}    & \textbf{3.07}    \\
    \hline
    \rowcolor{lightgray!40}
    \cellcolor[gray]{1} 
    & \#3  & PE                   & 7            & 267.17      & 95.881   & 11.41       & 95.86     & 95.90  & 95.884        & 2.07    & 2.05    \\
    \rowcolor{lightgray!40} 
    \cellcolor[gray]{1} 
    & TL   & APK$\rightarrow$PE & \textbf{4}            & \textbf{157.04}      & \textbf{96.034}   & \textbf{11.08}       & \textbf{95.95}     & \textbf{96.12}  & \textbf{96.035}        & \textbf{2.03}    & \textbf{1.94}    \\
    & \#3  & APK                  & 6            & 236.55      & 93.386   & \textbf{8.71}        & 93.73     & 93.04  & 93.412        & 3.11    & 3.48    \\
    \multirow{-4}{*}{75} 
    & TL   & PE$\rightarrow$APK & \textbf{3}            & \textbf{109.62}      & \textbf{93.927}   & 8.93        & \textbf{94.13}     & \textbf{93.72}  & \textbf{93.936}        & \textbf{2.92}    & \textbf{3.14}    \\
    \hline
    \rowcolor{lightgray!40}
    \cellcolor[gray]{1} 
    & \#3  & PE                   & 8            & 211.61      & 95.029   & 10.00       & 94.66     & 95.40  & 95.007        & 2.69    & 2.30    \\
    \rowcolor{lightgray!40} 
    \cellcolor[gray]{1} 
    & TL   & APK$\rightarrow$PE & \textbf{6}            & \textbf{141.66}      & \textbf{95.885}   & \textbf{9.97}        & \textbf{95.99}     & \textbf{95.78}  & \textbf{95.889}        & \textbf{2.00}    & \textbf{2.11} \\
    & \#3  & APK                  & 8            & 202.84      & 93.166   & 8.44        & 92.54     & \textbf{93.80}  & 93.121        & 3.78    & \textbf{3.10}    \\
    \multirow{-4}{*}{50} 
    & TL   & PE$\rightarrow$APK & \textbf{4}            & \textbf{114.17}      & \textbf{93.761}   & \textbf{8.34}        & \textbf{94.04}     & 93.48  & \textbf{93.781}        & \textbf{2.96}    & 3.26    \\
    \hline
    \rowcolor{lightgray!40}
    \cellcolor[gray]{1} 
    & \#3  & PE                   & 7            & 64.63       & 93.922   & 9.51        & 93.74     & 94.10  & 93.909        & 3.14    & 2.95    \\
    \rowcolor{lightgray!40} 
    \cellcolor[gray]{1} 
    & TL   & APK$\rightarrow$PE & \textbf{5}            & \textbf{33.31}       & \textbf{94.763}   & \textbf{9.12}        & \textbf{94.35}     & \textbf{95.18}  & \textbf{94.742}        & \textbf{2.85}    & \textbf{2.41}    \\
    & \#3  & APK                  & 6            & 44.09       & 91.584   & 9.04        & 91.33     & 91.84  & 91.563        & 4.36    & 4.08    \\
    \multirow{-4}{*}{10} 
    & TL   & PE$\rightarrow$APK & \textbf{5}            & \textbf{27.72}       & \textbf{92.737}   & \textbf{9.03}        & \textbf{92.65}     & \textbf{92.82}  & \textbf{92.735}        & \textbf{3.68}    & \textbf{3.59}    \\
    \hline
    \rowcolor{lightgray!40}
    \cellcolor[gray]{1} 
    & \#3  & PE                   & 7            & 41.99       & 93.276   & 12.36       & 93.47     & 93.08  & 93.295        & 3.25    & 3.46    \\
    \rowcolor{lightgray!40}
    \cellcolor[gray]{1} 
    & TL   & APK$\rightarrow$PE & \textbf{6}            & \textbf{22.21}       & \textbf{94.618}   & \textbf{10.49}       & \textbf{94.66}     & \textbf{94.58}  & \textbf{94.616}        & \textbf{2.67}    & \textbf{2.71}    \\
    & \#3  & APK                  & 8            & 37.54       & 91.189   & \textbf{7.74}        & 91.62     & 90.76  & 91.233        & 4.15    & 4.62    \\
    \multirow{-4}{*}{5} 
    & TL   & PE$\rightarrow$APK & \textbf{5}            & \textbf{18.44}       & \textbf{92.551}   & 7.80        & \textbf{92.78}     & \textbf{92.32}  & \textbf{92.567}        & \textbf{3.59}    & \textbf{3.84}  \\
  \bottomrule
\end{tabular}
\end{table*}

\begin{figure}[ht!]
  \centering
  \includegraphics[scale=0.6]{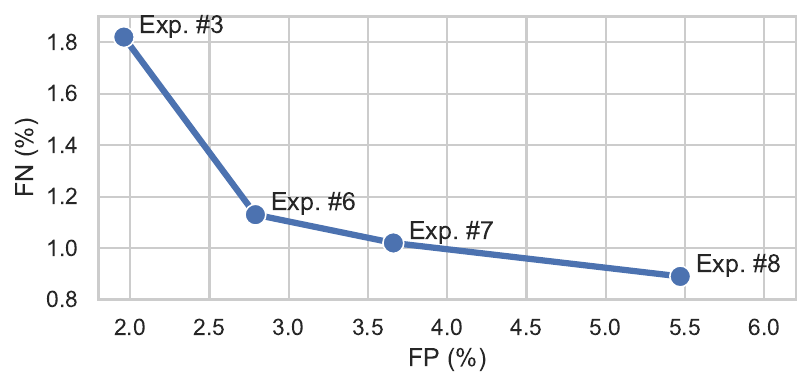}
  \caption{The trade-offs in FP/FN based on the relative cost assigned to each error (denoted by experiment number)} 
  \label{fig:sensativity_analysis_relation}
\end{figure}

\subsection{Analysis: Prioritizing Error Types}
\label{subsec:sensitivity}
 
Up to this point, the reward functions in all of our experiments assigned the same ``penalty'' to both error types: false positive and false negative. However, in many cases organizations may prefer to reduce one type of error at the expense of the other; organizations whose main priority is to protect their data may place greater emphasis on reducing false negatives, while other organizations may prefer to reduce the rate of their false alarms (i.e., false positives) so that it is within the capacity of human experts.

Our goal in this section is to analyze \MethodName's ability to adapt to scenarios such as those described above and to show that simple modifications to the reward function can realize the desired organizational policy. 
As a test case, we created three new variants of experiment \#3 on the PE dataset where the penalty for mistakenly classifying a malware as benign (i.e., false negative) is 10/50/100 times more expensive than that of classifying a benign file as malicious (i.e., false alarm). We denote these experiments as experiments \#6, \#7, and \#8 respectively.

The evaluation results are presented in Table~\ref{tab:sensativity_reward_setup} and Figure~\ref{fig:sensativity_analysis_relation}. It is clear that the modifications to the reward function achieve their stated goal and cause \MethodName to put greater emphasis on the prevention of false negatives. As shown in Table~\ref{tab:expactiondist_sensativity_pe}, the focus on the prevention of FNs is also reflected by \MethodName's choice of detectors: the more emphasis the reward function places on the prevention of FNs relative to the cost of FPs and computing resources, the more we use the computationally-expensive detectors.

The experiments in this section further show the effectiveness of using DRL to manage organizational security policy. When modeling malware detection as a classification problem, practitioners have only one tool at their disposal to shape their system's performance: the confidence score of their system (i.e., the score $0\leq x \leq 1$ produced by the classification algorithm). The confidence score can sometimes be a blunt instrument, partially because the desired FP/FN trade-offs need to be defined indirectly. \MethodName's reward function, on the other hand, enables us to explicitly define the relative priority we assign to each error type. Moreover, it enables us to easily define the desired efficiency of our model, an ability lacking in existing malware detection methods.

Finally, in order to demonstrate that \MethodName outperforms all baselines regarless of the relative weight they assign to different error types (i.e., FP vs. FN), we re-ran all baselines with different classification thresholds. In the additional experiments we varied the confidence threshold required to classify a file as malicious, setting it to the range of [0.3,0.7]. The full results, shown in Appendix~\ref{apndx:influence_threshold}, clearly show that \MethodName offers the best performance regardless of baseline configuration.

\subsection{Transfer Learning}
\label{subsec:transfer}

A major challenge in the field of information security is the deployment of a security solution in a new domain/organization. 
In addition to differences in the analyzed data and the data's characteristics, this task is challenging for two reasons: first, there is often a need to deploy the solution quickly, thus making long training times problematic. Secondly, there is often a shortage in labeled data that can be used to train the model.

In this section, we evaluate \MethodName's effectiveness in a \textit{transfer learning} (TL) scenario, where we take fully trained DRL architectures that were trained on one domain and then use them as the starting point for training on another domain. 
Our hypotheses were that such an approach can both reduce the required training time on the target dataset and improve performance when the amount of training data is limited in the target dataset.

To test our hypotheses, we conducted the following experiment: we extracted a fully trained DRL agent from each of our two datasets and applied it to the other (i.e., target) dataset. 
We tested various configurations for the target datasets, ranging from 100\% of the original training set to only 5\%. We used a standard technique in the application of transfer learning~\cite{zhu2011heterogeneous, shin2016deep}, where we reset the hyperparameters of the output layer and ``freeze'' all other layers (i.e., hard weight sharing). We used the reward function of experiment \#3 in all of our runs, since this configuration yielded excellent performance/efficiency trade-offs.

Our results are presented in Table~\ref{tab:transfer_analysis}, which shows the number of epochs required for the DRL agent to converge, the training duration in seconds, and performance statistics which include the F1 score and mean time per file. We draw two important conclusions:
\begin{itemize}
    \item \textbf{Significantly reduced training times.} Regardless of configuration, the use of TL reduces the time needed for training. The reduction is expressed both in the number of epochs and the overall training time, with the training time reduction ranging between 33-48\% for PE and 37-54\% for APK. Please note that we refer to the \textit{training time} (third column in Table~\ref{tab:transfer_analysis}) and not to the average runtime per file.
    \item \textbf{Improved performance when the target data is limited.} In all cases where the training set of the target dataset is reduced, \textit{the use of transfer learning improves model performance}. The improvement is evident in all metrics: F1, precision, recall, and accuracy.
\end{itemize}

\noindent Our experiments lead us to conclude that TL can potentially enable quick and effective deployment of \MethodName in new domains, even when the amount of available data is limited. It is also worth emphasizing that while both of our datasets are from the field of malware detection, each leverages a separate set of detectors with its own unique detection rates and score distributions (see Figure \ref{fig:toolsdist}). This means that in order to perform well, our DRL agent had to ``generalize'' the cost-benefit analysis presented earlier in this study.

\subsection{Adapting to Changes}
\label{subsec:adapting}

Throughout their lives, malware detection systems often have to adapt to changing circumstances: new types of attacks may make some detectors useless, vendors may terminate support for certain detectors, and new detectors may be added. 
To evaluate \MethodName's ability to quickly adapt to such changes, we created and tested three scenarios where the performance of one of our detectors -- \textit{pefile} -- is significantly altered. 
The scenarios were:

\begin{itemize}
    \item perfect model - 100\% \textit{correct} and confident predictions (either zero or one)
    \item inverse model - 100\% \textit{incorrect} and confident predictions (either zero or one)
    \item neutral model - provides only a single value of 0.5 (i.e., neutral) for all files
\end{itemize}

\noindent In the first two scenarios, the optimal course of action for \MethodName would be to rely completely on pefile (100\% incorrect predictions are just as useful as 100\% correct ones). 
In the third scenario, the optimal course of action would be to ignore pefile completely, since its predictions are random.

We used the fully trained policy generated in experiment \#3 and trained it for a single additional epoch. 
This short training achieved the desired results: for the first two scenarios, the model achieved an F1-score of 100\% (i.e., no mistakes). 
In the third scenario, \MethodName created a \textit{new action distribution that omits pefile completely}.

\begin{table}[ht]
    \small
	\caption{The action distribution of experiment \#3 after replacing \textit{pefile} with a neutral model.}
	\label{tab:expchange}
	\setlength\tabcolsep{3.7pt}
	\begin{tabular}{x{8.5mm}|cc|lcc}
		\toprule
		& F1  & Time & Action Sequences & Time & Files \\
		& (\%) & (sec) & & (sec) & (\%) \\
		\midrule
		\multirow{4}{*}{before}
		& \multirow{4}{*}{96.225} & \multirow{4}{*}{10.53}
         & byte3g                          &  3.99 & 83.38 \\
        &&& byte3g,pefile,opcode2g          & 48.98 & 12.67 \\
        &&& byte3g,pefile                   &  4.69 &  2.15 \\
        &&& byte3g,pefile,opcode2g,manalyze & 49.73 &  1.80 \\
		\hline
		\multirow{3}{*}{after}
        & \multirow{3}{*}{95.774} & \multirow{3}{*}{13.05}  
          & byte3g                      &  3.99 & 79.63 \\
        &&& byte3g,opcode2g             & 48.28 & 15.54 \\
        &&& byte3g,opcode2g,manalyze    & 49.03 &  4.83 \\
        \bottomrule
	\end{tabular}
\end{table}

This distribution is presented in Table~\ref{tab:expchange}. Both performance and runtime are slightly worse than the original experiment \#3 results due to the missing detector. 
We conclude that \MethodName can easily and quickly adapt to changes in detector performance.

\section{CONCLUSIONS}
In this study we present \MethodName, an RL-based approach for malware detection. 
Our approach dynamically assigns a subset of the available detectors to each file, constantly performing cost-benefit analysis to determine whether the use of a given detector is worth the expected reduction in classification uncertainty. 
The process is governed by a reward function which sets the relative rewards/costs of correct and incorrect classification, computational resources, runtime etc.
\MethodName has two main advantages.
First, it is highly efficient, since easy to classify files are likely to be assigned to a small set of efficient classifiers. 
This fact enables our method to maintain near optimal performance at a fraction of the computing cost.
Secondly, organizations can clearly and easily factor the costs of various resources (e.g., runtime) into \MethodName's decision-making process. Organizations are therefore able to define a security policy that best fits their threat exposure and budget. This approach is more expressive and refined than the standard use of confidence score threshold used by most classification algorithms.

In future work, we intend to tackle two advanced scenarios. First, we would like to reduce the costs by allowing parallel execution of detectors in the environment.
This use case can exponentially increase the action space, making it more challenging.
Secondly, we would like to extend our solution to support the ability to handle multiple files simultaneously. This requires the use in multi-agent architectures that share their resources.

\bibliographystyle{ACM-Reference-Format}
\bibliography{references}

\newpage
\appendix
\section{Further Analysis}

This appendix contains the complete results of our experiments and analysis, some of which we had to present in a condensed form in our study due to space constraints. We also present additional analysis, consisting of other evaluation metrics and hyperparameter configurations.

\subsection{Detector Analysis: Additional Thresholds}
\label{apndx:compensate_fp}

Our results in Table \ref{tab:compensate}, which showed that no detector dominates another in either of our two datasets, were calculated using the default detector setting. This means that for confidence scores greater than 0.5, the analyzed file was flagged malicious, and benign otherwise.
To further demonstrate that no detector is dominated by another, we repeated the experiments presented in Table~\ref{tab:compensate}, but with a fixed false positives rate 5\%. We argue that this scenario is typical of actual use cases where organizations determine a fixed false positives rate, setting its value at a level that would enable human experts to analyze all alerts.

The results of our additional experiments are presented in Table~\ref{tab:compensate_5}. The results show that for both datasets -- PE and APK -- no detector is dominated by another. Moreover, the large variance in the detection rates for the originally misclassified files -- larger than those presented in Table~\ref{tab:compensate} in some cases -- further supports our claim that an intelligent selection of detector subsets can maintain high performance at a fraction of the resources-use..

\begin{table}[htp!]
    \centering
    \caption{The performance of each detector on files that were misclassified by another detector when the FP rate is fixed at 5\%, for both datasets. 
    The results clearly show that no detector is dominated by another.}
    \label{tab:compensate_5}
    \renewcommand{\arraystretch}{0.9}
    \begin{subtable}{0.47\textwidth}
        \centering
    	\caption{PE dataset}
    	\setlength\tabcolsep{8.3pt}
    	\begin{tabular}{c|cccc}
    		\toprule
		            & manalyze & pefile  & byte3g  & opcode2g\\
    		\midrule
    		manalyze & -        & 82.18\% & 89.77\% & 91.01\% \\
    		pefile   & 67.37\%  & -       & 72.21\% & 77.18\% \\
    		byte3g   & 71.12\%  & 57.17\% & -       & 51.27\% \\
    		opcode2g & 73.09\%  & 62.70\% & 48.31\% & -       \\
    		\bottomrule
    	\end{tabular}
    \end{subtable}%
    \bigskip 
    \vspace*{2px}
    \begin{subtable}{0.47\textwidth}
        \centering
    	\caption{APK dataset}
    	\setlength\tabcolsep{8.1pt}
    	\begin{tabular}{c|cccc}
    		\toprule
    		          & manifest &  mmda   & bytecode & dalvikapi \\
    		\midrule
    		manifest  & -        & 48.43\% & 74.72\%  & 75.69\%   \\
    		mmda      & 29.39\%  & -       & 62.88\%  & 64.21\%   \\
    		bytecode  & 55.93\%  & 52.74\% & -        & 55.74\%   \\
    		dalvikapi & 41.07\%  & 36.63\% & 38.44\%  & -         \\
    		\bottomrule
    	\end{tabular}
    \end{subtable}
\end{table}

\newpage

\subsection{Influence of Changes in the Classification Threshold on Baseline Performance}
\label{apndx:influence_threshold}
The baseline F1-scores presented in Figure \ref{fig:pareto} were obtained using a confidence score threshold (i.e., the confidence level required by the baseline to classify a file as malicious) of 0.5. In order to show that \MethodName outperforms all baselines regardless of their confidence threshold settings, we conducted an additional set of experiments. 

\begin{figure}[ht]
    \centering
    \begin{subfigure}{.47\textwidth}
      \centering
      \includegraphics[scale=0.38]{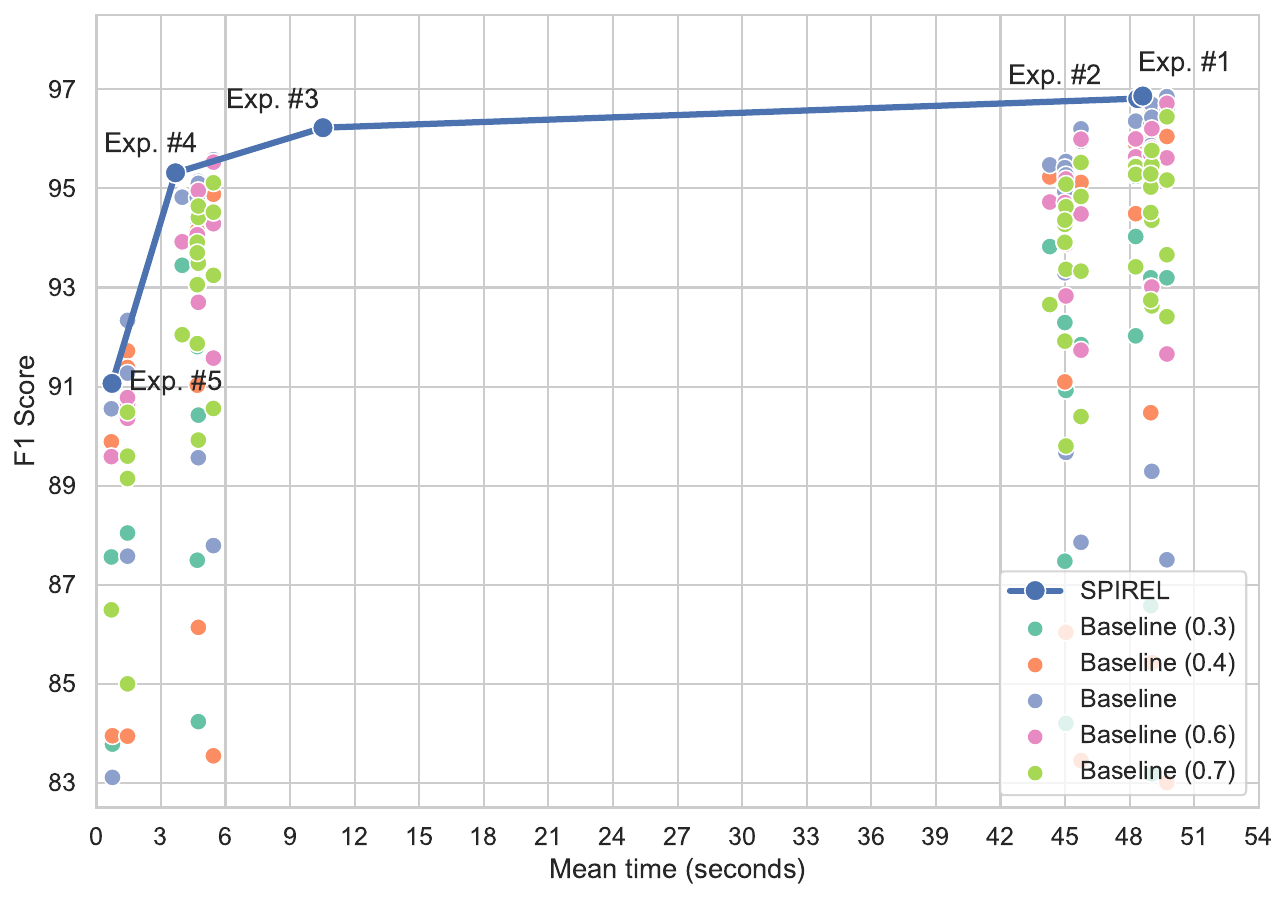}
      \caption{PE experiments}
      \vspace{2mm}%
    \end{subfigure}
    \begin{subfigure}{.47\textwidth}
      \centering
      \includegraphics[scale=0.38]{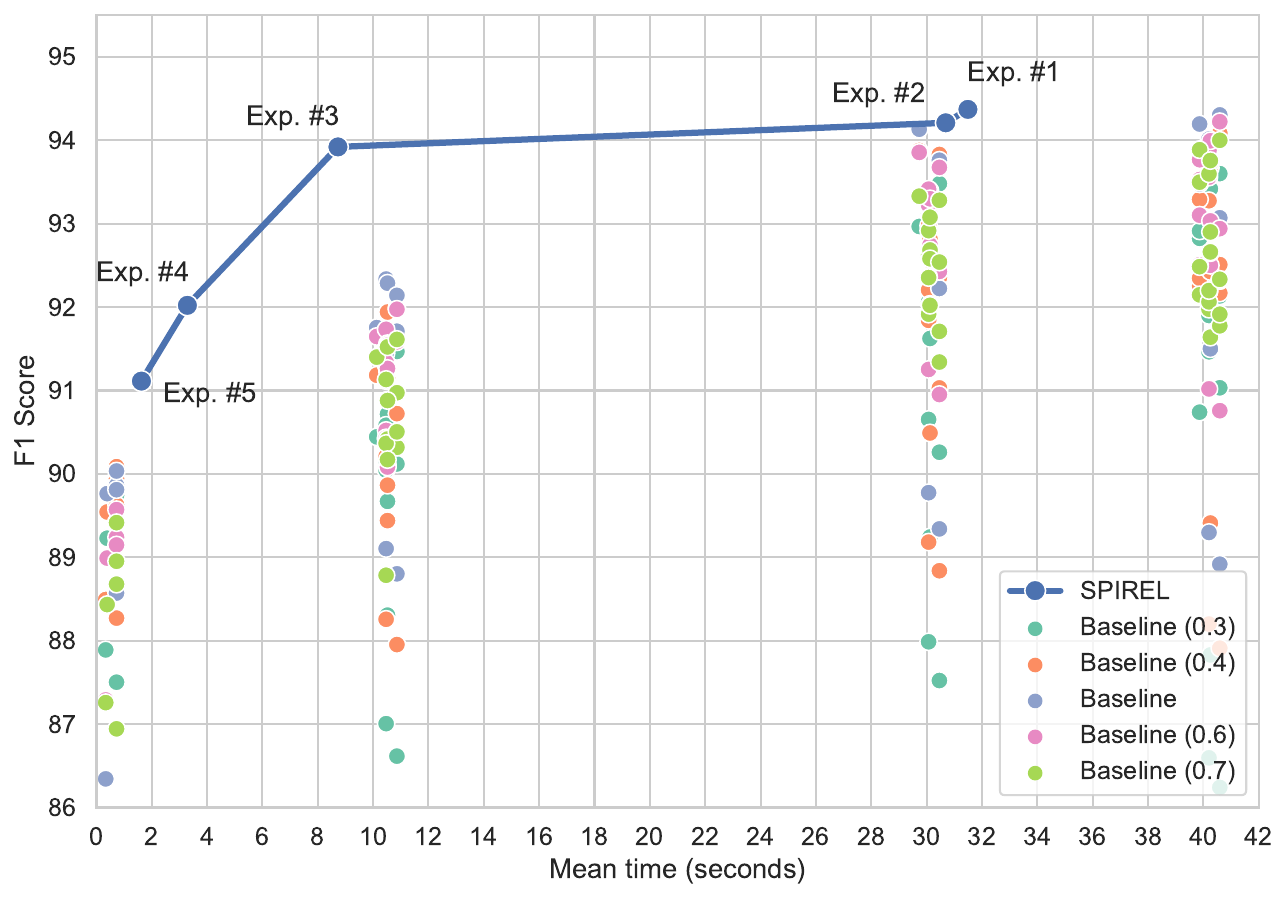}
      \caption{APK experiments}
    \end{subfigure}
  \caption{Sensitivity analysis of the influence of different classification thresholds (from 0.3 to 0.7) over all baselines, compared to \MethodName.} 
  \label{fig:sensativity_analysis_threshold_changes}
\end{figure}

Figure \ref{fig:sensativity_analysis_threshold_changes} presents our additional experiments. We ran each baseline in four additional configurations, with the confidence score threshold set to different values. The chosen threshold values were set to 0.3,0.4, 0.6, and 0.7 (0.5 is the original value presented in Figure \ref{fig:pareto}). Our results clearly show that \MethodName outperforms all baselines, regardless of their chosen configurations.

\newpage

\subsection{Using Accuracy as the Evaluation Metric}
\label{apndx:pareto_acc}
Throughout this study, we have used F1-score as the main evaluation metric. While we have also reported the results of the accuracy metric, space constraints did not permit us to plot the results on a graph in order to determine whether \MethodName forms the Pareto Frontier for this metric as well.

\begin{figure}[htp!]
    \centering
    \begin{subfigure}{.47\textwidth}
      \centering
      \includegraphics[scale=0.46]{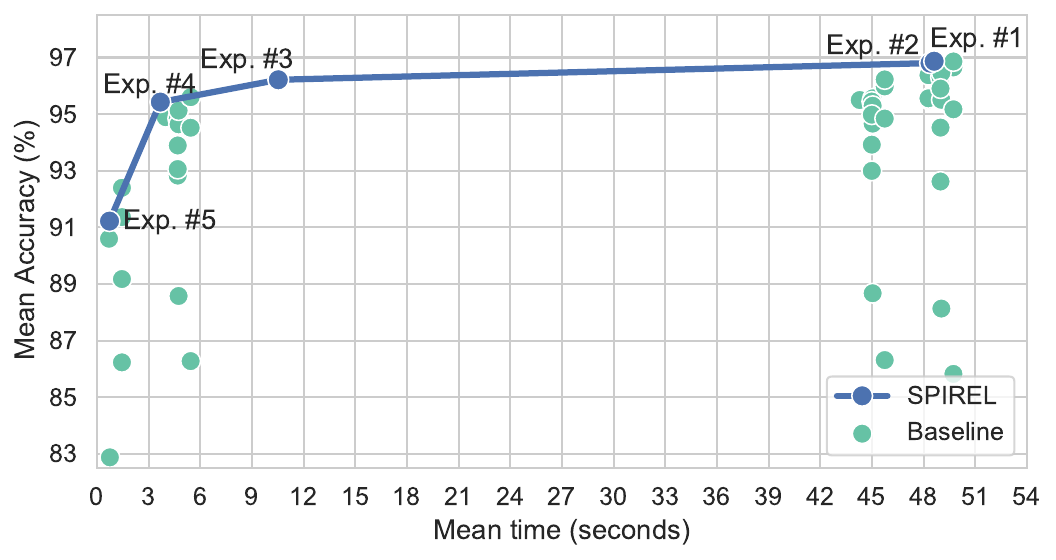}
      \caption{PE dataset}
      \vspace{2mm}
    \end{subfigure}
    \begin{subfigure}{.47\textwidth}
      \centering
      \includegraphics[scale=0.46]{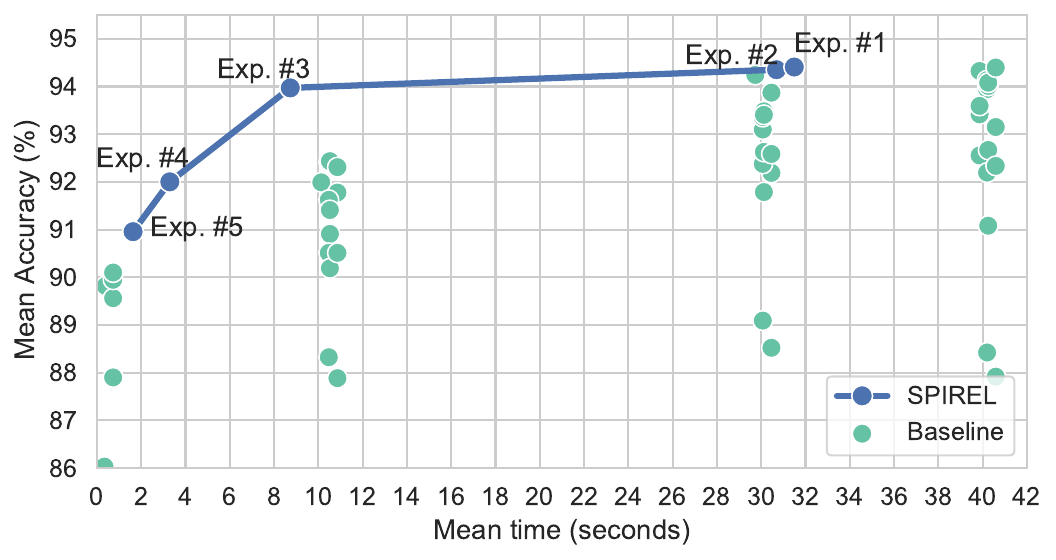}
      \caption{APK dataset}
    \end{subfigure}
  \caption{The results of the experiments conducted on each dataset, compared to the results of their corresponding baselines, with regard to the accuracy and mean time per file.} 
  \label{fig:pareto_acc}
\end{figure}

Figure~\ref{fig:pareto_acc} presents a comparison of the mean accuracy and time of all baselines and \MethodName experiments, for both datasets. The graphs clearly show that \MethodName once again outperforms all baselines. Additionally, it is clear that \MethodName once again forms the Pareto frontier for both datasets.

\newpage

\subsection{DRL Agent Action Distribution}
\label{apndx:drl_agent_action_distribution}

In continuation to the analysis presented in subsection~\ref{subsec:analysis}, we now present in full detail the detector combinations chosen by each \MethodName experiment. The results are presented in 
Table~\ref{tab:expactiondist} and include the description of the detector subset, its average runtime and the percentage of files (out of all those analyzed) on which this particular detectors subset was applied.

\begin{table}[htp!]
    \small
    \centering
	\caption{Distribution of detector combination choices made by the agent for each of our experimental policies. 
	The detector combinations are sorted in descending order according to their percentage of use for each experiment. The corresponding mean time per file is also presented.
	}
	\label{tab:expactiondist}
	\renewcommand{\arraystretch}{0.95}
	\setlength\tabcolsep{4.5pt}
	\begin{tabular}{c|c|Hlcc}
		\toprule
		\multicolumn{2}{c|}{Exper. \#} & Acc. (\%) & Action Sequences & Time (sec) & Files (\%)\\
		\midrule
		\multirow{16}{*}{PE}
        & \multirow{3}{*}{1} & \multirow{3}{*}{96.867}
          & manalyze,opcode2g,byte3g,pefile & 49.73 & 66.76 \\
        &&& manalyze,opcode2g,pefile        & 45.74 & 28.02 \\
        &&& manalyze,opcode2g,pefile,byte3g & 49.73 &  5.22 \\
        \cmidrule{2-6}
        & \multirow{3}{*}{2} & \multirow{3}{*}{96.801}
          & opcode2g,manalyze,pefile,byte3g & 49.73 & 63.54 \\
        &&& opcode2g,pefile                 & 44.99 & 27.29 \\
        &&& opcode2g,pefile,byte3g          & 48.98 & 9.17 \\
        \cmidrule{2-6}
        & \multirow{4}{*}{3} & \multirow{4}{*}{96.212}
          & byte3g                          &  3.99 & 83.38 \\
        &&& byte3g,pefile,opcode2g          & 48.98 & 12.67 \\
        &&& byte3g,pefile                   &  4.69 &  2.15 \\
        &&& byte3g,pefile,opcode2g,manalyze & 49.73 &  1.80 \\
        \cmidrule{2-6}
        & \multirow{4}{*}{4} & \multirow{4}{*}{95.424}
          & manalyze,byte3g,pefile          &  5.44 & 50.77 \\
        &&& manalyze,pefile                 &  1.45 & 22.49 \\
        &&& manalyze                        &  0.75 & 16.89 \\
        &&& manalyze,byte3g                 &  4.74 &  9.85 \\
        \cmidrule{2-6}
        & \multirow{2}{*}{5} & \multirow{2}{*}{91.220}
          & pefile                          &  0.70 & 96.17 \\
        &&& pefile,manalyze                 &  1.45 &  3.83 \\
        
        \midrule
        
		\multirow{27}{*}{APK}
        & \multirow{4}{*}{1} & \multirow{4}{*}{94.409}
          & dalvikapi                        & 29.73 & 83.82 \\
        &&& dalvikapi,bytecode,manifest,mmda & 40.59 & 11.18 \\
        &&& dalvikapi,bytecode               & 39.86 &  3.23 \\
        &&& dalvikapi,bytecode,manifest      & 40.20 &  1.77 \\
        \cmidrule{2-6}
        & \multirow{5}{*}{2} & \multirow{5}{*}{94.353}
          & dalvikapi                        & 29.73 & 85.94 \\
        &&& dalvikapi,mmda,bytecode,manifest & 40.59 &  5.25 \\
        &&& dalvikapi,manifest,bytecode,mmda & 40.59 &  3.41 \\
        &&& dalvikapi,manifest               & 30.07 &  3.02 \\
        &&& dalvikapi,mmda                   & 30.12 &  2.38 \\
        \cmidrule{2-6}
        & \multirow{6}{*}{3} & \multirow{6}{*}{93.971}
          & mmda,bytecode                    & 10.52 & 43.89 \\
        &&& mmda                             &  0.39 & 43.23 \\
        &&& mmda,bytecode,dalvikapi,manifest & 40.59 &  4.35 \\
        &&& mmda,bytecode,manifest           & 10.86 &  4.31 \\
        &&& mmda,bytecode,dalvikapi          & 40.25 &  2.15 \\
        &&& mmda,bytecode,manifest,dalvikapi & 40.59 &  2.07 \\
        \cmidrule{2-6}
        & \multirow{7}{*}{4} & \multirow{7}{*}{92.001}
           & mmda                             &  0.39 & 90.35 \\
        &&& mmda,bytecode                    & 10.52 &  2.67 \\
        &&& mmda,bytecode,dalvikapi          & 40.25 &  2.10 \\
        &&& mmda,bytecode,dalvikapi,manifest & 40.59 &  1.75 \\
        &&& mmda,dalvikapi                   & 30.12 &  1.67 \\
        &&& mmda,dalvikapi,bytecode          & 40.25 &  0.93 \\
        &&& mmda,dalvikapi,bytecode,manifest & 40.59 &  0.53 \\
        \cmidrule{2-6}
        & \multirow{5}{*}{5} & \multirow{5}{*}{90.956}
           & mmda                             &  0.39 & 93.95 \\
         &&& mmda,bytecode,manifest           & 10.86 &  2.34 \\
         &&& mmda,bytecode                    & 10.52 &  1.63 \\
         &&& mmda,bytecode,dalvikapi,manifest & 40.59 &  1.12 \\
         &&& mmda,bytecode,dalvikapi          & 40.25 &  0.96 \\
        \bottomrule
	\end{tabular}
\end{table}
\newpage

\subsection{Training Times}
\label{apndx:training_times}
The training time required for our DRL-agent to convergence change according to experimental settings. The experiments conducted in this study had the same action space and state representation, and a varying reward function (i.e., confusion matrix) that indicates the relation between correct and incorrect classification. Table \ref{tab:training_times} specifies for every dataset the amount of epochs and the average total time required for it to convergence. 

\begin{table}[htp!]
    \centering
	\caption{The training duration and number of epochs required to converge for both datasets. } 
	\label{tab:training_times}
	\renewcommand{\arraystretch}{0.93}
	\setlength\tabcolsep{11pt}
	\begin{tabular}{c|c|cc}
		\toprule
		\multicolumn{1}{c|}{}& Experiment (\#) & Epochs (\#) & Time (min.) \\
		\midrule
		\multirow{5}{*}{PE} 
		& 1  & 6 &  285.62 \\
		& 2    & 6 &  278.32\\
		& 3    & 5 &  242.96 \\
		& 4  & 6 & 265.49  \\
		& 5  & 5 & 245.06 \\
		\midrule
    	\multirow{5}{*}{APK} 
		& 1  & 5 &  210.71 \\
		& 2  & 4 &  195.93 \\
		& 3  & 4 & 187.45 \\
		& 4 & 6 & 252.68 \\
		& 5 & 4 & 181.80 \\
		\bottomrule
	\end{tabular}
\end{table}

\subsection{Experimental Results}
\label{apndx:full_baselines}
As mentioned in Section~\ref{sec:introduction}, there is no existing ML approach for the dynamic selection of detector subsets while a file is being analyzed. 
For this reason, any baseline we compare our approach to must consist of a preselected detector subset (i.e., defined prior to running). 
We therefore compare our method to all possible detector subset combinations and their aggregation methods. 

Given that we have four detectors and three aggregation methods (or, majority, and stacking), we evaluated 48 baselines for each of our datasets.
Tables \ref{tab:combinations_pe_full} and \ref{tab:combinations_apk_full} present the F1 score and average runtime for all runs. 
An analysis of the results shows that \MethodName consistently outperforms all of the baselines. 
More specifically, for every desired level of performance (with the exception of the poorest performing baselines which consist of a single detector), there is a version of our proposed method that can result in the equivalent or better performance at a reduced runtime. 
\MethodName’s superior performance is particularly evident when compared to the top performing baselines. 
For both datasets, our approach outperformed the top performing baselines while being more computationally efficient.

\begin{table*}[htp!]
    \small
    \centering
	\caption{Performance and mean time values of all possible detector combinations for PE files. 
	The 10-fold cross-validated baseline results are sorted in descending order by their mean F1 score and are presented along with the results of the experiments described in Section~\ref{subsec:experimental_results}.
	Aggregations were conducted using: \textit{majority}, \textit{or}, \textit{stacking} using the \textit{Decision Tree and Random Forest} methods (see Section~\ref{subsec:performance}). 
    }
	\label{tab:combinations_pe_full}
	\renewcommand{\arraystretch}{1.05}
	\setlength\tabcolsep{5.6pt}
	\begin{tabular}{clcccccccc}
		\toprule
		\# & Detector Combination & Aggregation & F1 score (\%) & Time (sec) & Precision (\%) & Recall (\%) & Accuracy (\%) & FP (\%) & FN (\%) \\
		\midrule
        \rowcolor{lightgray!40} - &
        Experiment \#1                   &     SPIREL     & 96.865 &  48.61 & 97.23 & 96.51 & 96.867 &  1.38 &  1.75 \\
        \newtag{1}{bl_pe} &
        manalyze,pefile,byte3g,opcode2g  & stacking (RF)  & 96.852 &  49.73 & 96.94 & 96.76 & 96.859 &  1.52 &  1.62 \\
        \rowcolor{lightgray!40} - &
        Experiment \#2                   &     SPIREL     & 96.812 &  48.37 & 96.45 & 97.18 & 96.801 &  1.79 &  1.41 \\
        \newtag{2}{bl_pe} &
        manalyze,byte3g,opcode2g         &    majority    & 96.693 &  49.03 & 97.07 & 96.32 & 96.709 &  1.45 &  1.84 \\
        \newtag{3}{bl_pe} &
        manalyze,pefile,byte3g,opcode2g  &    majority    & 96.626 &  49.73 & 97.17 & 96.09 & 96.649 &  1.40 &  1.95 \\
        \newtag{4}{bl_pe} &
        manalyze,byte3g,opcode2g         & stacking (RF)  & 96.436 &  49.03 & 96.83 & 96.04 & 96.455 &  1.57 &  1.98 \\
        \newtag{5}{bl_pe} &
        byte3g,opcode2g                  &    majority    & 96.358 &  48.28 & 96.68 & 96.04 & 96.374 &  1.65 &  1.98 \\
        \newtag{6}{bl_pe} &
        pefile,byte3g,opcode2g           &    majority    & 96.279 &  48.98 & 96.74 & 95.82 & 96.301 &  1.61 &  2.09 \\
        \rowcolor{lightgray!40} - &
        Experiment \#3                   &     SPIREL     & 96.225 &  10.53 & 96.09 & 96.36 & 96.212 &  1.96 &  1.82 \\
        \newtag{7}{bl_pe} &
        manalyze,pefile,opcode2g         & stacking (RF)  & 96.202 &  45.74 & 96.66 & 95.75 & 96.224 &  1.65 &  2.12 \\
        \newtag{8}{bl_pe} &
        manalyze,pefile,opcode2g         &    majority    & 95.953 &  45.74 & 96.42 & 95.49 & 95.978 &  1.77 &  2.25 \\
        \newtag{9}{bl_pe} &
        pefile,byte3g,opcode2g           & stacking (RF)  & 95.866 &  48.98 & 96.57 & 95.18 & 95.901 &  1.69 &  2.41 \\
        \newtag{10}{bl_pe} &
        byte3g,opcode2g                  &       or       & 95.652 &  48.28 & 93.79 & 97.59 & 95.569 &  3.23 &  1.20 \\
        \newtag{11}{bl_pe} &
        manalyze,pefile,byte3g           &    majority    & 95.591 &   5.44 & 96.06 & 95.13 & 95.618 &  1.95 &  2.43 \\
        \newtag{12}{bl_pe} &
        manalyze,pefile,byte3g           & stacking (RF)  & 95.577 &   5.44 & 96.08 & 95.08 & 95.599 &  1.94 &  2.46 \\
        \newtag{13}{bl_pe} &
        manalyze,opcode2g                &    majority    & 95.543 &  45.04 & 95.73 & 95.35 & 95.557 &  2.12 &  2.32 \\
        \newtag{14}{bl_pe} &
        byte3g,opcode2g                  & stacking (RF)  & 95.543 &  48.28 & 95.73 & 95.35 & 95.557 &  2.12 &  2.32 \\
        \newtag{15}{bl_pe} &
        manalyze,byte3g,opcode2g         & stacking (DT)  & 95.495 &  49.03 & 95.58 & 95.41 & 95.505 &  2.20 &  2.29 \\
        \newtag{16}{bl_pe} &
        opcode2g                         &      none      & 95.475 &  44.29 & 95.83 & 95.13 & 95.497 &  2.07 &  2.43 \\
        \newtag{17}{bl_pe} &
        byte3g,opcode2g                  & stacking (DT)  & 95.470 &  48.28 & 96.26 & 94.69 & 95.513 &  1.84 &  2.65 \\
        \newtag{18}{bl_pe} &
        pefile,opcode2g                  &    majority    & 95.419 &  44.99 & 95.84 & 95.01 & 95.444 &  2.06 &  2.49 \\
        \rowcolor{lightgray!40} - &
        Experiment \#4                   &     SPIREL     & 95.315 &   3.68 & 97.77 & 92.98 & 95.424 &  1.06 &  3.51 \\
        \newtag{19}{bl_pe} &
        manalyze,opcode2g                & stacking (RF)  & 95.273 &  45.04 & 95.60 & 94.95 & 95.294 &  2.18 &  2.52 \\
        \newtag{20}{bl_pe} &
        manalyze,pefile,byte3g,opcode2g  & stacking (DT)  & 95.169 &  49.73 & 95.07 & 95.27 & 95.169 &  2.47 &  2.36 \\
        \newtag{21}{bl_pe} &
        manalyze,byte3g                  &    majority    & 95.143 &   4.74 & 95.14 & 95.14 & 95.149 &  2.43 &  2.43 \\
        \newtag{22}{bl_pe} &
        manalyze,byte3g                  & stacking (RF)  & 95.102 &   4.74 & 95.44 & 94.77 & 95.125 &  2.26 &  2.61 \\
        \newtag{23}{bl_pe} &
        pefile,opcode2g                  & stacking (RF)  & 94.942 &  44.99 & 95.53 & 94.36 & 94.979 &  2.20 &  2.82 \\
        \newtag{24}{bl_pe} &
        manalyze,pefile,opcode2g         & stacking (DT)  & 94.837 &  45.74 & 94.81 & 94.87 & 94.842 &  2.60 &  2.56 \\
        \newtag{25}{bl_pe} &
        byte3g                           &      none      & 94.823 &   3.99 & 95.98 & 93.69 & 94.890 &  1.96 &  3.15 \\
        \newtag{26}{bl_pe} &
        pefile,byte3g                    &    majority    & 94.813 &   4.69 & 95.30 & 94.33 & 94.846 &  2.32 &  2.83 \\
        \newtag{27}{bl_pe} &
        manalyze,opcode2g                & stacking (DT)  & 94.653 &  45.04 & 94.74 & 94.57 & 94.664 &  2.62 &  2.71 \\
        \newtag{28}{bl_pe} &
        manalyze,byte3g                  & stacking (DT)  & 94.604 &   4.74 & 95.13 & 94.08 & 94.640 &  2.41 &  2.96 \\
        \newtag{29}{bl_pe} &
        manalyze,pefile,byte3g           & stacking (DT)  & 94.520 &   5.44 & 94.38 & 94.67 & 94.518 &  2.82 &  2.66 \\
        \newtag{30}{bl_pe} &
        pefile,byte3g,opcode2g           & stacking (DT)  & 94.513 &  48.98 & 94.57 & 94.46 & 94.522 &  2.71 &  2.77 \\
        \newtag{31}{bl_pe} &
        pefile,opcode2g                  & stacking (DT)  & 93.911 &  44.99 & 93.93 & 93.89 & 93.920 &  3.03 &  3.05 \\
        \newtag{32}{bl_pe} &
        pefile,byte3g                    & stacking (RF)  & 93.848 &   4.69 & 94.41 & 93.29 & 93.892 &  2.76 &  3.35 \\
        \newtag{33}{bl_pe} &
        pefile,opcode2g                  &       or       & 93.296 &  44.99 & 89.35 & 97.61 & 92.994 &  5.81 &  1.19 \\
        \newtag{34}{bl_pe} &
        pefile,byte3g                    &       or       & 93.091 &   4.69 & 89.70 & 96.75 & 92.829 &  5.55 &  1.63 \\
        \newtag{35}{bl_pe} &
        pefile,byte3g                    & stacking (DT)  & 93.057 &   4.69 & 92.97 & 93.14 & 93.059 &  3.52 &  3.42 \\
        \newtag{36}{bl_pe} &
        pefile,byte3g,opcode2g           &       or       & 93.013 &  48.98 & 88.19 & 98.39 & 92.618 &  6.58 &  0.80 \\
        \newtag{37}{bl_pe} &
        manalyze,pefile                  &    majority    & 92.336 &   1.45 & 92.96 & 91.72 & 92.396 &  3.47 &  4.14 \\
        \newtag{38}{bl_pe} &
        manalyze,pefile                  & stacking (RF)  & 91.272 &   1.45 & 92.11 & 90.45 & 91.361 &  3.87 &  4.77 \\
        \rowcolor{lightgray!40} - &
        Experiment \#5                   &     SPIREL     & 91.065 &   0.73 & 92.71 & 89.48 & 91.220 &  3.52 &  5.26 \\
        \newtag{39}{bl_pe} &
        pefile                           &      none      & 90.552 &   0.70 & 90.88 & 90.23 & 90.597 &  4.52 &  4.88 \\
        \newtag{40}{bl_pe} &
        manalyze,opcode2g                &       or       & 89.672 &  45.04 & 82.32 & 98.46 & 88.673 & 10.56 &  0.77 \\
        \newtag{41}{bl_pe} &
        manalyze,byte3g                  &       or       & 89.565 &   4.74 & 82.35 & 98.17 & 88.576 & 10.51 &  0.91 \\
        \newtag{42}{bl_pe} &
        manalyze,byte3g,opcode2g         &       or       & 89.289 &  49.03 & 81.27 & 99.06 & 88.131 & 11.40 &  0.47 \\
        \newtag{43}{bl_pe} &
        manalyze,pefile                  & stacking (DT)  & 89.145 &   1.45 & 89.28 & 89.01 & 89.174 &  5.34 &  5.49 \\
        \newtag{44}{bl_pe} &
        manalyze,pefile,opcode2g         &       or       & 87.858 &  45.74 & 78.87 & 99.16 & 86.312 & 13.27 &  0.42 \\
        \newtag{45}{bl_pe} &
        manalyze,pefile,byte3g           &       or       & 87.791 &   5.44 & 78.99 & 98.80 & 86.276 & 13.13 &  0.60 \\
        \newtag{46}{bl_pe} &
        manalyze,pefile                  &       or       & 87.578 &   1.45 & 79.70 & 97.18 & 86.231 & 12.36 &  1.41 \\
        \newtag{47}{bl_pe} &
        manalyze,pefile,byte3g,opcode2g  &       or       & 87.505 &  49.73 & 78.15 & 99.40 & 85.823 & 13.88 &  0.30 \\
        \newtag{48}{bl_pe} &
        manalyze                         &      none      & 83.113 &   0.75 & 81.89 & 84.38 & 82.876 &  9.32 &  7.80 \\
        \bottomrule
	\end{tabular}
\end{table*}

\begin{table*}[htp!]
    \small
    \centering
	\caption{Performance and mean time values of all possible detectors combinations for APK files. 
	The 10-fold cross-validated baseline results are sorted in descending order by their mean F1-score and are presented along with the results of the experiments described in Section~\ref{subsec:experimental_results}.
	Aggregations were conducted using: \textit{majority}, \textit{or}, \textit{stacking} using \textit{Decision Tree and Random Forest} methods (see~\ref{subsec:performance}). 
    }
	\label{tab:combinations_apk_full}
    \renewcommand{\arraystretch}{1.05}
	\setlength\tabcolsep{5.3pt}
	\begin{tabular}{clcccccccc}
		\toprule
		\# & Detector Combination & Aggregation & F1 score (\%) & Time (sec) & Precision (\%) & Recall (\%) & Accuracy (\%) & FP (\%) & FN (\%) \\
		\midrule
        \rowcolor{lightgray!40} - &
        Experiment \#1                   &     SPIREL     & 94.369 &  31.49 & 95.07 & 93.68 & 94.409 &  2.43 &  3.16 \\
        \newtag{1}{bl_apk} &
        manifest,mmda,bytecode,dalvikapi & stacking (RF)  & 94.303 &  40.59 & 96.04 & 92.63 & 94.399 &  1.91 &  3.69 \\
        \rowcolor{lightgray!40} - &
        Experiment \#2                   &     SPIREL     & 94.208 &  30.69 & 96.64 & 91.90 & 94.353 &  1.60 &  4.05 \\
        \newtag{2}{bl_apk} &
        bytecode,dalvikapi               &    majority    & 94.194 &  39.86 & 96.53 & 91.97 & 94.326 &  1.66 &  4.02 \\
        \newtag{3}{bl_apk} &
        dalvikapi                        &      none      & 94.131 &  29.73 & 96.07 & 92.27 & 94.242 &  1.89 &  3.87 \\
        \newtag{4}{bl_apk} &
        manifest,bytecode,dalvikapi      & stacking (RF)  & 94.013 &  40.20 & 96.06 & 92.05 & 94.133 &  1.89 &  3.98 \\
        \newtag{5}{bl_apk} &
        mmda,bytecode,dalvikapi          & stacking (RF)  & 93.975 &  40.25 & 95.76 & 92.25 & 94.080 &  2.04 &  3.88 \\
        \rowcolor{lightgray!40} - &
        Experiment \#3                   &     SPIREL     & 93.919 &   8.73 & 94.24 & 93.60 & 93.971 &  2.86 &  3.20 \\
        \newtag{6}{bl_apk} &
        mmda,bytecode,dalvikapi          &    majority    & 93.880 &  40.25 & 96.10 & 91.76 & 94.013 &  1.87 &  4.12 \\
        \newtag{7}{bl_apk} &
        manifest,bytecode,dalvikapi      &    majority    & 93.849 &  40.20 & 95.44 & 92.31 & 93.945 &  2.20 &  3.85 \\
        \newtag{8}{bl_apk} &
        manifest,mmda,dalvikapi          & stacking (RF)  & 93.760 &  30.46 & 95.59 & 92.00 & 93.871 &  2.13 &  4.00 \\
        \newtag{9}{bl_apk} &
        bytecode,dalvikapi               &       or       & 93.479 &  39.86 & 92.68 & 94.29 & 93.417 &  3.73 &  2.86 \\
        \newtag{10}{bl_apk} &
        bytecode,dalvikapi               & stacking (RF)  & 93.476 &  39.86 & 95.32 & 91.70 & 93.595 &  2.25 &  4.15 \\
        \newtag{11}{bl_apk} &
        mmda,dalvikapi                   &    majority    & 93.395 &  30.12 & 94.70 & 92.13 & 93.480 &  2.58 &  3.94 \\
        \newtag{12}{bl_apk} &
        mmda,dalvikapi                   & stacking (RF)  & 93.323 &  30.12 & 94.60 & 92.08 & 93.406 &  2.63 &  3.97 \\
        \newtag{13}{bl_apk} &
        manifest,dalvikapi               & stacking (RF)  & 93.253 &  30.07 & 94.86 & 91.70 & 93.359 &  2.49 &  4.15 \\
        \newtag{14}{bl_apk} &
        manifest,dalvikapi               &    majority    & 93.116 &  30.07 & 92.96 & 93.28 & 93.098 &  3.54 &  3.36 \\
        \newtag{15}{bl_apk} &
        manifest,mmda,bytecode,dalvikapi &    majority    & 93.070 &  40.59 & 94.26 & 91.91 & 93.150 &  2.80 &  4.05 \\
        \newtag{16}{bl_apk} &
        mmda,bytecode,dalvikapi          & stacking (DT)  & 92.652 &  40.25 & 92.96 & 92.35 & 92.670 &  3.50 &  3.83 \\
        \newtag{17}{bl_apk} &
        mmda,dalvikapi                   & stacking (DT)  & 92.595 &  30.12 & 93.09 & 92.11 & 92.628 &  3.42 &  3.95 \\
        \newtag{18}{bl_apk} &
        manifest,mmda,dalvikapi          & stacking (DT)  & 92.547 &  30.46 & 93.12 & 91.98 & 92.586 &  3.40 &  4.01 \\
        \newtag{19}{bl_apk} &
        bytecode,dalvikapi               & stacking (DT)  & 92.507 &  39.86 & 93.18 & 91.85 & 92.555 &  3.36 &  4.08 \\
        \newtag{20}{bl_apk} &
        manifest,dalvikapi               & stacking (DT)  & 92.357 &  30.07 & 92.68 & 92.03 & 92.377 &  3.64 &  3.99 \\
        \newtag{21}{bl_apk} &
        manifest,bytecode                &    majority    & 92.335 &  10.47 & 93.38 & 91.31 & 92.414 &  3.24 &  4.35 \\
        \newtag{22}{bl_apk} &
        manifest,mmda,bytecode,dalvikapi & stacking (DT)  & 92.318 &  40.59 & 92.60 & 92.03 & 92.335 &  3.68 &  3.99 \\
        \newtag{23}{bl_apk} &
        mmda,bytecode                    &    majority    & 92.286 &  10.52 & 94.21 & 90.44 & 92.435 &  2.78 &  4.79 \\
        \newtag{24}{bl_apk} &
        manifest,mmda,dalvikapi          &    majority    & 92.223 &  30.46 & 91.90 & 92.55 & 92.189 &  4.08 &  3.73 \\
        \newtag{25}{bl_apk} &
        manifest,bytecode,dalvikapi      & stacking (DT)  & 92.188 &  40.20 & 92.34 & 92.03 & 92.194 &  3.82 &  3.99 \\
        \newtag{26}{bl_apk} &
        manifest,mmda,bytecode           & stacking (RF)  & 92.140 &  10.86 & 94.30 & 90.08 & 92.309 &  2.73 &  4.96 \\
        \newtag{27}{bl_apk} &
        mmda,dalvikapi                   &       or       & 92.033 &  30.12 & 89.42 & 94.80 & 91.787 &  5.61 &  2.60 \\
        \rowcolor{lightgray!40} - &
        Experiment \#4                   &     SPIREL     & 92.021 &   3.29 & 91.78 & 92.26 & 92.001 &  4.13 &  3.87 \\
        \newtag{28}{bl_apk} &
        bytecode                         &      none      & 91.754 &  10.13 & 94.63 & 89.05 & 91.991 &  2.53 &  5.48 \\
        \newtag{29}{bl_apk} &
        manifest,mmda,bytecode           &    majority    & 91.711 &  10.86 & 92.53 & 90.91 & 91.776 &  3.67 &  4.55 \\
        \newtag{30}{bl_apk} &
        mmda,bytecode,dalvikapi          &       or       & 91.500 &  40.25 & 87.46 & 95.93 & 91.082 &  6.88 &  2.04 \\
        \newtag{31}{bl_apk} &
        manifest,bytecode                & stacking (RF)  & 91.447 &  10.47 & 93.52 & 89.47 & 91.625 &  3.10 &  5.27 \\
        \newtag{32}{bl_apk} &
        mmda,bytecode                    & stacking (RF)  & 91.252 &  10.52 & 93.05 & 89.52 & 91.411 &  3.34 &  5.25 \\
        \newtag{33}{bl_apk} &
        mmda,bytecode                    &       or       & 91.225 &  10.52 & 88.23 & 94.42 & 90.909 &  6.30 &  2.79 \\
        \rowcolor{lightgray!40} - &
        Experiment \#5                   &     SPIREL     & 91.113 &   1.63 & 89.60 & 92.68 & 90.956 &  5.38 &  3.66 \\
        \newtag{34}{bl_apk} &
        manifest,mmda,bytecode           & stacking (DT)  & 90.468 &  10.86 & 90.96 & 89.98 & 90.512 &  4.47 &  5.02 \\
        \newtag{35}{bl_apk} &
        manifest,bytecode                & stacking (DT)  & 90.418 &  10.47 & 91.40 & 89.46 & 90.512 &  4.21 &  5.28 \\
        \newtag{36}{bl_apk} &
        mmda,bytecode                    & stacking (DT)  & 90.107 &  10.52 & 90.93 & 89.30 & 90.188 &  4.46 &  5.36 \\
        \newtag{37}{bl_apk} &
        manifest,mmda                    & stacking (RF)  & 90.035 &   0.73 & 90.70 & 89.38 & 90.099 &  4.59 &  5.31 \\
        \newtag{38}{bl_apk} &
        manifest,mmda                    &    majority    & 89.863 &   0.73 & 87.41 & 92.46 & 89.561 &  6.67 &  3.77 \\
        \newtag{39}{bl_apk} &
        manifest,mmda                    & stacking (DT)  & 89.811 &   0.73 & 91.12 & 88.54 & 89.948 &  4.32 &  5.74 \\
        \newtag{40}{bl_apk} &
        manifest,dalvikapi               &       or       & 89.776 &  30.07 & 84.54 & 95.71 & 89.091 &  8.76 &  2.15 \\
        \newtag{41}{bl_apk} &
        mmda                             &      none      & 89.764 &   0.39 & 90.31 & 89.23 & 89.817 &  4.79 &  5.39 \\
        \newtag{42}{bl_apk} &
        manifest,mmda,dalvikapi          &       or       & 89.342 &  30.46 & 83.44 & 96.14 & 88.521 &  9.55 &  1.93 \\
        \newtag{43}{bl_apk} &
        manifest,bytecode,dalvikapi      &       or       & 89.299 &  40.20 & 83.07 & 96.53 & 88.422 &  9.84 &  1.73 \\
        \newtag{44}{bl_apk} &
        manifest,bytecode                &       or       & 89.106 &  10.47 & 83.57 & 95.43 & 88.323 &  9.39 &  2.29 \\
        \newtag{45}{bl_apk} &
        manifest,mmda,bytecode,dalvikapi &       or       & 88.919 &  40.59 & 82.16 & 96.89 & 87.915 & 10.53 &  1.56 \\
        \newtag{46}{bl_apk} &
        manifest,mmda,bytecode           &       or       & 88.803 &  10.86 & 82.60 & 96.01 & 87.884 & 10.12 &  2.00 \\
        \newtag{47}{bl_apk} &
        manifest,mmda                    &       or       & 88.573 &   0.73 & 83.97 & 93.71 & 87.900 &  8.96 &  3.15 \\
        \newtag{48}{bl_apk} &
        manifest                         &      none      & 86.346 &   0.34 & 84.50 & 88.28 & 86.029 &  8.10 &  5.87 \\
        \bottomrule
	\end{tabular}
\end{table*}

\end{document}